\documentclass[
  reprint,
  aps,
  prx,
  amsmath,
  amssymb,
  superscriptaddress,
  longbibliography
]{revtex4-2}

\usepackage{graphicx}
\usepackage{bbm}
\usepackage{subfigure}
\usepackage{braket}
\usepackage{hyperref}
\usepackage{xcolor}
\usepackage{comment}
\usepackage{hyperref}

\begin{document}

\title{High-order correlations and ultrafast Wigner negativities in bright-squeezed-vacuum-driven high-harmonic generation}
% Multi-mode
% Third-order correlation

\author{Sebasti\'an de-la-Pe\~na}
\email{sebastian.delapena@mpsd.mpg.de}
\affiliation{Max Planck Institute for the Structure and Dynamics of Matter, Luruper Chaussee 149, 22761 Hamburg, Germany}

\author{Heiko Appel}
\affiliation{Max Planck Institute for the Structure and Dynamics of Matter, Luruper Chaussee 149, 22761 Hamburg, Germany}

\author{Marcelo F. Ciappina}
\affiliation{Department of Physics, Guangdong Technion-Israel Institute of Technology, Shantou 515063, China}
\affiliation{Technion--Israel Institute of Technology, Haifa 32000, Israel}
\affiliation{Guangdong Provincial Key Laboratory of Materials and Technologies for Energy Conversion, Guangdong Technion-Israel Institute of Technology, Shantou 515063, China}

\author{Ofer Neufeld}
\email{ofern@technion.ac.il}
\affiliation{Technion - Israel Institute of Technology, Schulich Faculty of Chemistry, 3200003 Haifa, Israel}

\author{Angel Rubio}
\email{angel.rubio@mpsd.mpg.de}
\affiliation{Max Planck Institute for the Structure and Dynamics of Matter, Luruper Chaussee 149, 22761 Hamburg, Germany}
\affiliation{Initiative for Computational Catalysis (ICC), Flatiron Institute, 162 5th Ave, New York, NY 10010, USA}

\date{\today}
\maketitle

%%%%%%%%%%%%%%%%%%%%%%%%%%%%%%%%%%%%%%%%%%%%%%%%%%%%%%%%%%%%%%%%%%
%                            Abstract                            %
%%%%%%%%%%%%%%%%%%%%%%%%%%%%%%%%%%%%%%%%%%%%%%%%%%%%%%%%%%%%%%%%%%

\onecolumngrid
\section*{Abstract}

High-harmonic generation (HHG) is a prototypical strong-field process in which intense light drives matter to emit radiation at integer multiples of the driving frequency. Extending HHG into the quantum-optical regime offers new opportunities to probe and control strongly nonlinear light–matter interactions using nonclassical states of light. Yet describing this regime requires a fully quantum treatment of the correlated electron–photon dynamics, which becomes computationally challenging for broadband, strongly squeezed fields. Here we solve the quantum-electrodynamical dynamics of a two-level system driven by bright squeezed vacuum in a converged multimode Hilbert space. Both the driving field and emitted harmonics are fully quantized, with the light–matter interaction treated nonperturbatively. This enables direct access to the multimode quantum state and its higher-order correlations beyond semiclassical sampling or perturbative descriptions. We show that squeezed-vacuum driving produces harmonic emission with qualitatively distinct second- and third-order photon correlations compared with coherent excitation. Moreover, back-action from the driven emitter strongly reshapes the incident squeezed field, generating pronounced Wigner-function negativities that evolve on attosecond timescales. Our results establish a fully quantum framework for broadband strong-field dynamics with squeezed light and provide a route to predicting and interpreting quantum-HHG experiments and their extension to more complex emitters.

\vspace{0.5em}

%\twocolumngrid

%%%%%%%%%%%%%%%%%%%%%%%%%%%%%%%%%%%%%%%%%%%%%%%%%%%%%%%%%%%%%%%%%%
%                        Introduction                            %
%%%%%%%%%%%%%%%%%%%%%%%%%%%%%%%%%%%%%%%%%%%%%%%%%%%%%%%%%%%%%%%%%%

%\section*{}

\section{Introduction}

%% HISTORY OF HHG

High-harmonic generation (HHG) is a strongly nonlinear process in which a low-frequency laser interacts with a material sample, be it gas~\cite{Ferray1988, McPherson1987}, liquid~\cite{Luu2018, Mondal2023}, or solid~\cite{Ghimire2010}; resulting in the emission of higher harmonics of the incident field. HHG has enabled the development of new research areas, most notably attosecond metrology~\cite{Hentschel2001} and attosecond spectroscopy~\cite{Krausz2009}. The theoretical description of HHG was first formulated within a semiclassical framework for the electron dynamics and the incident light, known as the three-step model~\cite{Corkum1993}. This picture was subsequently developed using a quantum-mechanical description of the electron~\cite{Lewenstein1994, Antoine1996}, while the light field continued to be described within classical electromagnetic theory, thereby neglecting quantum-optical effects. \\

%% QHHG CONTEXT

\noindent However, recent works suggest that quantum-optical effects can be prominent in HHG under certain conditions, either in the emitted harmonics~\cite{Gonoskov2016, Tsatrafyllis2017, Lewenstein2021, Gorlach2020, Gombkt2021, Stammer2022, Stammer2023, Pizzi2023, Stammer2024, Theidel2024, Theidel2025, delaPea2025} or in the radiated field~\cite{Tzur2022, EvenTzur2023, Gorlach2023, Tzur2024, EvenTzur2024-2, delaPea2024, Rasputnyi2024, Tzur2025, Yi2025, RiveraDean2025, Klimkin2025, Liu2026, RiveraDean2026, Gonoskov2026}. Particular interest has been raised by the case of bright squeezed vacuum (BSV), in which the radiated field has a vanishing average electric field amplitude at all times, while containing a nonzero number of photons and non-vanishing variance. Despite its recent use in both theoretical~\cite{Gorlach2023, EvenTzur2024-2, Liu2026, RiveraDean2026} and experimental~\cite{Rasputnyi2024} studies, an open question remains as to whether BSV can give rise to quantum effects beyond the mere sampling of a probability distribution of light, i.e. truly quantum features such as entanglement and negative Wigner distributions. More specifically, although several approaches have been used to model HHG driven by nonclassical light, they treat the driving field as an ensemble of classical fields sampled from an appropriate distribution, and can therefore capture effects arising from the statistical properties of the field. However, such descriptions do not retain the full quantum correlations between the light and matter~\cite{delaPea2024}, and it therefore remains unclear whether the multimode quantum correlations of BSV give rise to HHG effects that cannot be reproduced by a semiclassical ensemble, particularly in the infinite-photon limit. \\

%% OUR WORK
%% REQUIRES MAJOR REVISION

\begin{figure*}[t]
\centering
\includegraphics[width=0.6\linewidth]{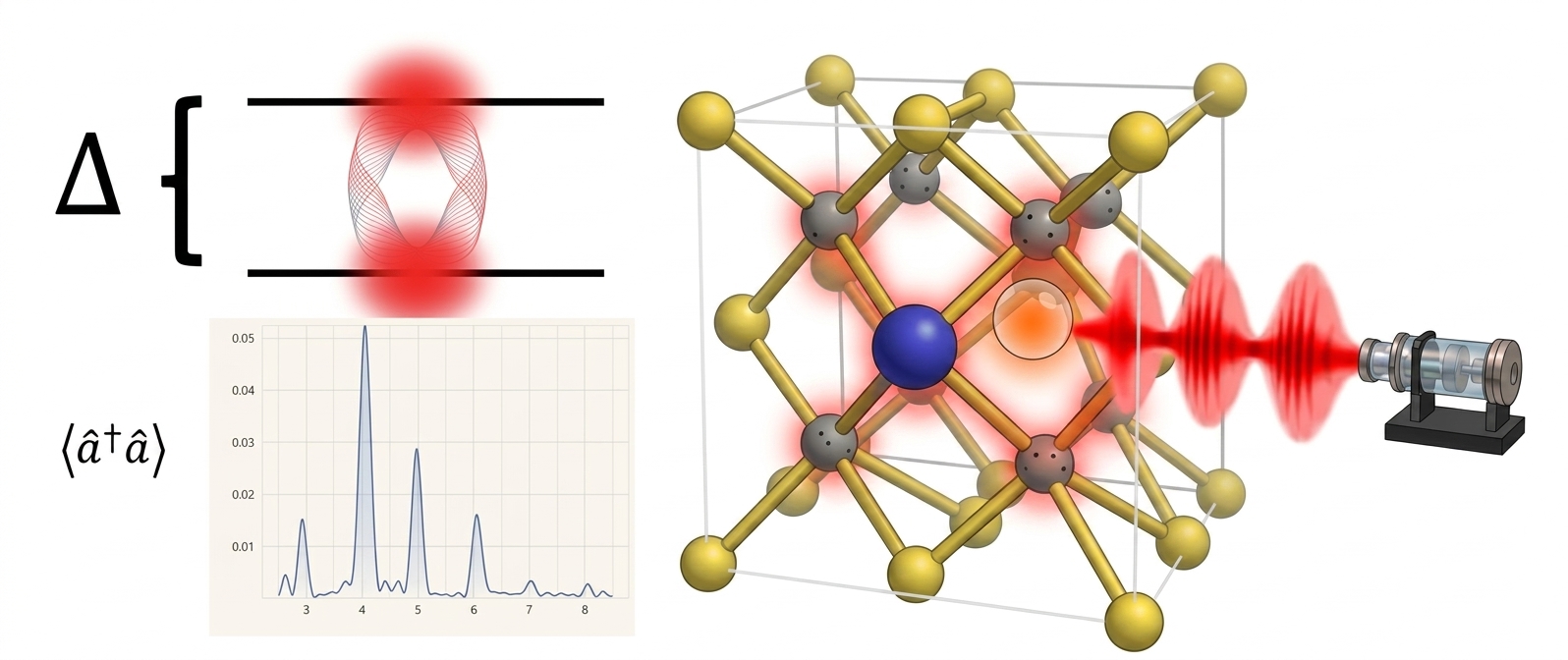}
\caption{Schematic depiction of a diamond nano-vacancy center (modelled by a two-level system) driven by a bright-squeezed laser pulse, depicted as oscillations of electric field quadratures, and emitting higher harmonics.}
\label{fig: depiction}
\end{figure*}

\noindent Here, we propagate the quantum-electrodynamical (QED) Hamiltonian of a two-level system (TLS) driven by a BSV mode. The TLS is chosen as the simplest setting in which the theory can be benchmarked against an exact solution. At the same time, it can be viewed as a minimal two-band model of a solid near the $\Gamma$ point, where the dynamics are restricted to a valence and a conduction state coupled by the driving field. Such reduced two-band descriptions are widely used to capture the essential interband dynamics underlying strong-field excitation and high-harmonic generation in solids. The model may also represent genuinely localized two-level systems, such as selected transitions in diamond NV centers~\cite{Barhum2026} (see illustration in Fig.~\ref{fig: depiction}). We calculate the HHG emission spectrum and compare it with that obtained under classical coherent driving, as well as second- and third-order photon correlations. To propagate the equations of motion, we employ a squeezed picture for the driving mode~\cite{EvenTzur2024-2}, which allows us to reach the infinite-photon-number limit of BSV while retaining the self-consistent light-matter interaction throughout the propagation. This allows directly addressing the above questions since quantum optical correlations are kept in full. As the system evolves, we find that the Wigner distribution of the pump develops clear negative regions, with dynamics evolving on attosecond timescales. This provides a direct signature of the genuinely quantum back-action induced by BSV driving and motivates experimental approaches capable of probing these effects. More broadly, our framework treats quantized light and matter dynamics on equal footing, enabling the study of back-action on quantum driving fields, collective strong coupling in multi-emitter systems, and higher-order correlations between harmonic modes.

%% REQUIRES MAJOR REVISION

%%%%%%%%%%%%%%%%%%%%%%%%%%%%%%%%%%%%%%%%%%%%%%%%%%%%%%%%%%%%%%%%%%
%                        Theoretical model                            %
%%%%%%%%%%%%%%%%%%%%%%%%%%%%%%%%%%%%%%%%%%%%%%%%%%%%%%%%%%%%%%%%%%

\section{Theoretical model} 

We begin describing the QED Hamiltonian in the interaction picture for a two-level system coupled to a driving mode and a set of continuum emission modes,

\begin{align}
    \hat{H}_{\mathrm{QED}}(t) = \hat{H}_{\mathrm{TLS}} + \hat{H}_{\mathrm{driv}}(t) + \hat{H}_{\mathrm{cont}}(t),
\end{align}

\noindent where $\hat{H}_{\mathrm{TLS}} = (1-\hat{\sigma}_z)\Delta $ is the TLS Hamiltonian with $\Delta$ being the energy gap, $\hat{\sigma}_z$ a Pauli matrix, and

\begin{align}
    \hat{H}_{\mathrm{driv}}(t) = \hat{\sigma}_x \lambda \sqrt{\frac{\omega_L}{2}}  \left[ \hat{a}_L e^{-i\omega_L t} + \hat{a}^\dagger_L e^{i\omega_L t} \right],
    \label{eq: driving-ham}
\end{align}

\begin{align}
    \hat{H}_{\mathrm{cont}}(t) = \hat{\sigma}_x  \sum_{\eta \in \mathcal{D}} \lambda\sqrt{\frac{\omega_\eta}{2}} \left[ \hat{a}_\eta e^{-i\omega_\eta t} + \hat{a}^\dagger_\eta e^{i\omega_\eta t} \right],
    \label{eq: continuum-ham}
\end{align}

%  (harmonics will be formally defined as bandwidths in $\eta$ around a specific integer)
\noindent are the driving-mode and the continuum-mode Hamiltonians, respectively. Here, $\omega_L$ is the driving frequency, while $\omega_\eta=\eta \omega_L$ is the frequency of the continuum mode labeled by $\eta$ denoting the harmonic label. The operators $\hat{a}_\eta^{(\dagger)}$ and $\hat{a}_L^{(\dagger)}$ are the annihilation (creation) operators for the continuum mode labeled by $\eta$ and the driving mode, respectively. The light-matter coupling is given by $\lambda = g \sqrt{\Delta \omega}$, where $g$ is the spectral light-matter coupling (see SI) and $\Delta \omega$ is frequency spacing. The sum over $\eta$ is restricted to the relevant spectral domain $\mathcal{D}$ for each particular calculation (see SI). The initial state of the driving mode is a BSV, $\ket{\xi}_L = \hat{S}(\xi) \ket{0}_L$, where $\xi$ is the squeezing parameter and $\hat{S}(\xi)$ is the squeezing operator. The continuum of modes and the TLS are initiated in their respective ground-states. \\

\noindent We perform the calculation such that the proper continuum limit $\Delta \omega \rightarrow 0$ is taken, which consequently implies the light-matter coupling to vanish, $\lambda \rightarrow 0$. This limit affects the driving and continuum Hamiltonians differently. For the continuum Hamiltonian, decreasing $\Delta\omega$ simply increases the density of modes included in the calculation. For the driving Hamiltonian, however, the situation is different because it contains only a single mode. To take the continuum limit in this case, we transform to the squeezed picture by defining a transformed light-matter wavefunction $\ket{\Psi^{(s)}(t)} = \hat{S}^\dagger(\xi) \ket{\Psi(t)}$~\cite{EvenTzur2024-2}, which leads to the transformation on operators $\hat{\mathcal{O}}^{(s)} = \hat{S}^\dagger(\xi) \hat{\mathcal{O}} \hat{S}(\xi)$. Consequently, the Hamiltonian in Eq.~\eqref{eq: driving-ham} becomes

\begin{align}
    \hat{H}^{(s)}_{\mathrm{driv}}(t) = \kappa  \hat{\sigma}_x \left[ \left( \hat{a}_L \cosh{\left|\xi\right|} - e^{i\varphi} \hat{a}^\dagger_L \sinh{\left|\xi\right|} \right) e^{-i\omega_L t} + \left( \hat{a}^\dagger_L \cosh{\left|\xi\right|} - e^{-i\varphi} \hat{a}_L \sinh{\left|\xi\right|} \right) e^{i\omega_L t} \right],
\label{eq: driving-ham-sq}
\end{align}

\noindent where the creation and annihilation operators $\hat{a}_L^{(\dagger)}$ now act on a squeezed-Fock Hilbert space rather than ordinary Fock Hilbert space, $\left| \xi \right|$ and $\varphi$ are the complex amplitude and phase of the initial squeezing parameter $\xi$, and $\kappa = \lambda\sqrt{\omega_L/2}$ simply agglutinates all prefactors for the driving coupling. Note that the transformation only affects the operators for the driving mode and leaves any of the continuum modes unchanged. Moreover, the advantage of this picture is that the initial state of light remains a (squeezed-photon) vacuum regardless of the initial squeezing $\xi$, $\ket{\Psi^{(s)}(t=0)} = \ket{0}_{\mathrm{TLS}}\otimes \ket{0}_L \otimes_{\eta} \ket{0}_\eta$, as the squeezing strength is now in the transformed Hamiltonian. \\

\noindent The transformed Hamiltonian of Eq.~\eqref{eq: driving-ham-sq} allows us to consider the physically relevant regime of a strongly squeezed driving while keeping the initial state fixed. In particular, the limit of large squeezing, $\left| \xi \right| \rightarrow \infty$~\cite{Gorlach2023, RiveraDean2026-2}, corresponds to an increasing number of photons in the initial state of the original unsqueezed representation. Simultaneously, the continuum limit can be taken, such that we make $\kappa \propto \sqrt{\Delta \omega } \rightarrow 0$. Taking simultaneously $\Delta\omega\rightarrow0$ and $|\xi|\rightarrow\infty$ in Eq.~\eqref{eq: driving-ham-sq} while retaining a finite driving strength $\sqrt{\Delta \omega} e^{\left| \xi \right|}$ gives

\begin{align}
    \hat{H}^{(s)}_{\mathrm{driv}}(t) \rightarrow \frac{\Omega_R}{2} \hat{\sigma}_x  \left[ \left( \hat{a}_L - e^{i\varphi} \hat{a}^\dagger_L \right) e^{-i\omega_L t} +  \left( \hat{a}^\dagger_L - e^{-i\varphi} \hat{a}_L \right) e^{i\omega_L t} \right],
\label{eq: driving-ham-bsv}
\end{align}

\noindent where all driving-intensity dependence is now captured by the bright-squeezed Rabi frequency $\Omega_R = \kappa e^{\left| \xi \right|}$. Equation~\eqref{eq: driving-ham-bsv} makes explicit why $\Omega_R$ is the relevant parameter in this limit: after taking $|\xi|\gg1$, the dependence on the initial squeezing and light-matter coupling appears only through the combination $\kappa e^{|\xi|}$. Consequently, different combinations of $\kappa$ and $\xi$ that give the same $\Omega_R$ lead to the same limiting driving Hamiltonian. The limit is formally exact within the $|\xi|\rightarrow\infty$ regime, since the terms proportional to $e^{-|\xi|}$ vanish in this limit. Although the full Hamiltonian from Eq.\eqref{eq: driving-ham-sq} is employed in the simulations, Eq.\eqref{eq: driving-ham-bsv} shows that, once the strongly squeezed regime is reached, the driving strength relevant to the TLS dynamics is characterized by $\Omega_R$, rather than by $\kappa$ and $\xi$ independently. Additionally, Eq.~\eqref{eq: driving-ham-bsv} demonstrates that the truncation of the squeezed-Fock Hilbert space required for convergence depends uniquely on $\Omega_R$, rather than on the squeezing parameter and light-matter coupling independently. Hence, we find that the limit $\Delta\omega\rightarrow 0$ can be taken provided that the mode is strongly squeezed ($\left| \xi \right| \gg 1$) and the initial squeezing is captured through the Rabi frequency ($\Omega_R \propto \sqrt{\Delta \omega} \exp{\left\lbrace \left| \xi \right| \right\rbrace}$). In other words, the continuum limit and infinite-photon limit can be taken jointly while keeping the physical driving strength finite, with the effect of the increasing photon number absorbed into $\Omega_R$.

%%%%%%%%%%%%%%%%%%%%%%%%%%%%%%%%%%%%%%%%%%%%%%%%%%%%%%%%%%%%%%%%%%
%                        Results and discussion                            %
%%%%%%%%%%%%%%%%%%%%%%%%%%%%%%%%%%%%%%%%%%%%%%%%%%%%%%%%%%%%%%%%%%

\section{Results and discussion}

Applying this approach, we take the driving frequency $\omega_L$ as the reference unit. The relevant physical parameters are therefore the gap/driving frequency ratio $\Delta/\omega_L$, and the Rabi/driving-frequency ratio $\Omega_R/\omega_L$. The squeezing phase is set to zero, $\varphi = 0$, such that the field variance is initially off: $\mu \sqrt{\bra{\xi}\hat{E}_{\mathrm{L}}^2(0) \ket{\xi}} = \Omega_R \left| \sin{(- \varphi/2)} \right| = 0$. When emission is concerned, it is also physically relevant to specify the ratio associated with the dimensionless spectral light-matter coupling $g = 2\mu \sqrt{\alpha/A_{\mathrm{eff}}}$ (see SI), where $\mu$ is the dipole transition amplitude between the two levels, $A_{\mathrm{eff}}$ is the effective area of the one-dimensional waveguide, and $\alpha$ is the fine-structure constant. Here, $\mu$ and $A_{\mathrm{eff}}$ are specified in physical atomic units ($g$ is dimensionless). We choose $A_{\mathrm{eff}} \approx 320~\mathrm{nm}^2$ and $\mu \approx 0.053~\mathrm{nm}$, which gives $g \approx 5\times10^{-4}$. The frequency spacing $\Delta\omega$ is formally converged in every simulation. We consider two representative solid-state-inspired driving scenarios. The first is motivated by GaAs driven at a wavelength of $800$ nm, which provides an example close to resonant interband excitation. The second is motivated by ZnO driven in the infrared, around $1900$ nm, representative of an off-resonant multiphoton regime. To study these limits systematically within the TLS, we consider cases characterized by the ratio between the level separation and the photon energy, $\Delta/\omega_L$. In particular, we analyze resonant driving with $\Delta/\omega_L=1$ and nonresonant driving with $\Delta/\omega_L=5$, approximately corresponding to the GaAs and ZnO examples, respectively. \\

\subsection{HHG emission and correlation spectra}

%% Explicit spectrum

\begin{figure*}[t]
\centering

\begin{minipage}{0.48\textwidth}
    \centering
    \includegraphics[width=\linewidth]{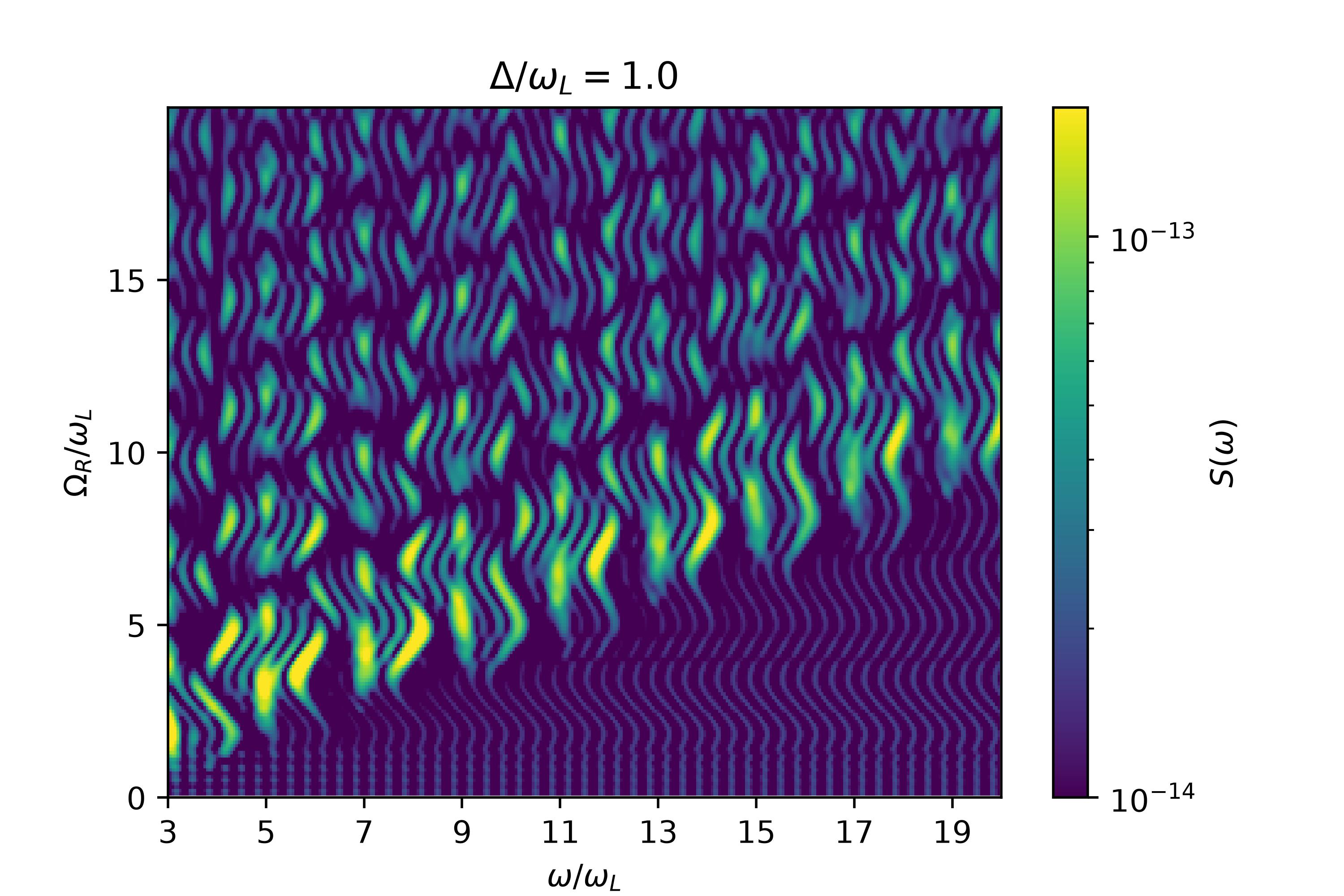}
\end{minipage}
\hfill
\begin{minipage}{0.48\textwidth}
    \centering
    \includegraphics[width=\linewidth]{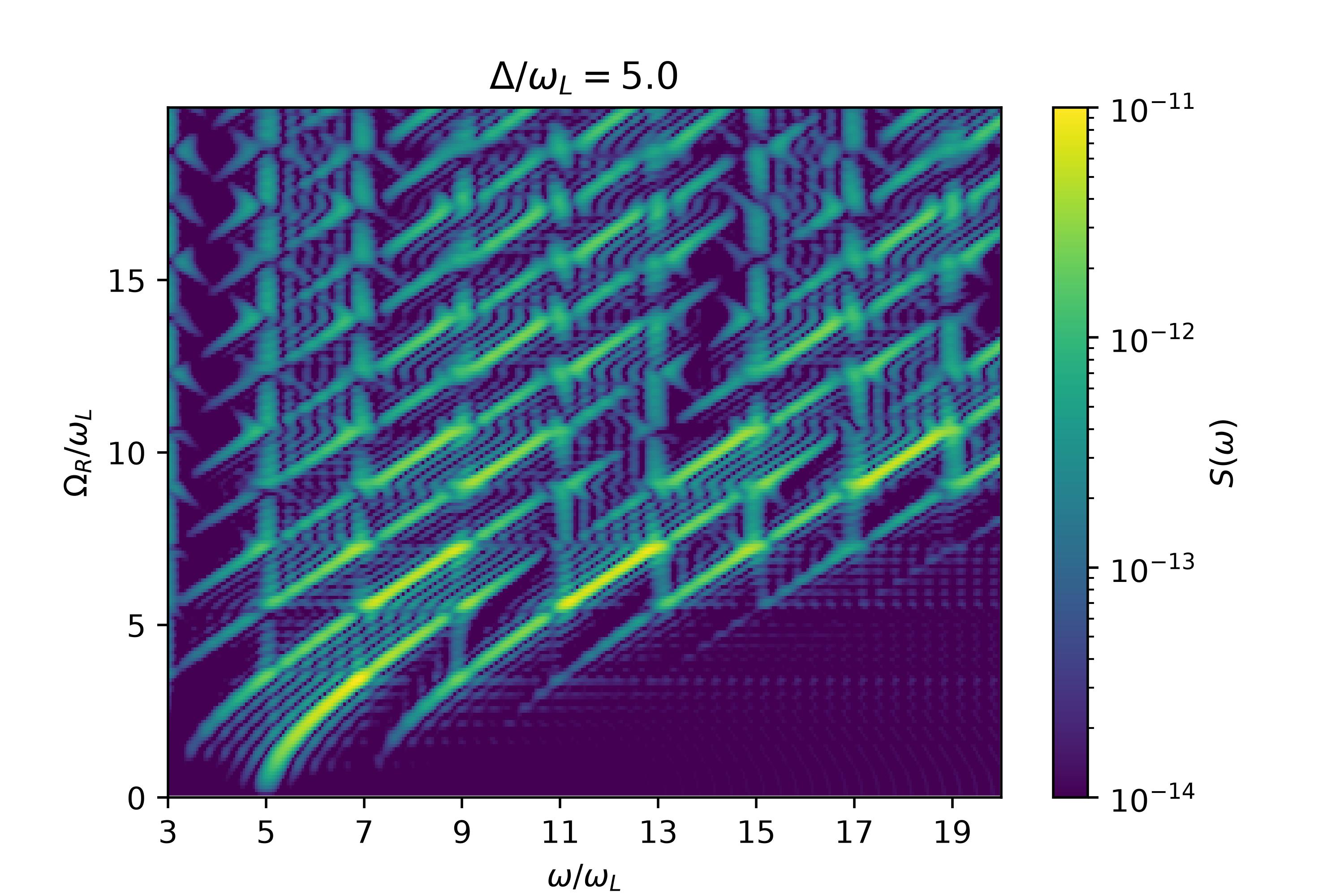}
\end{minipage}

\vspace{0.1cm}

\begin{minipage}{0.48\textwidth}
    \centering
    \includegraphics[width=\linewidth]{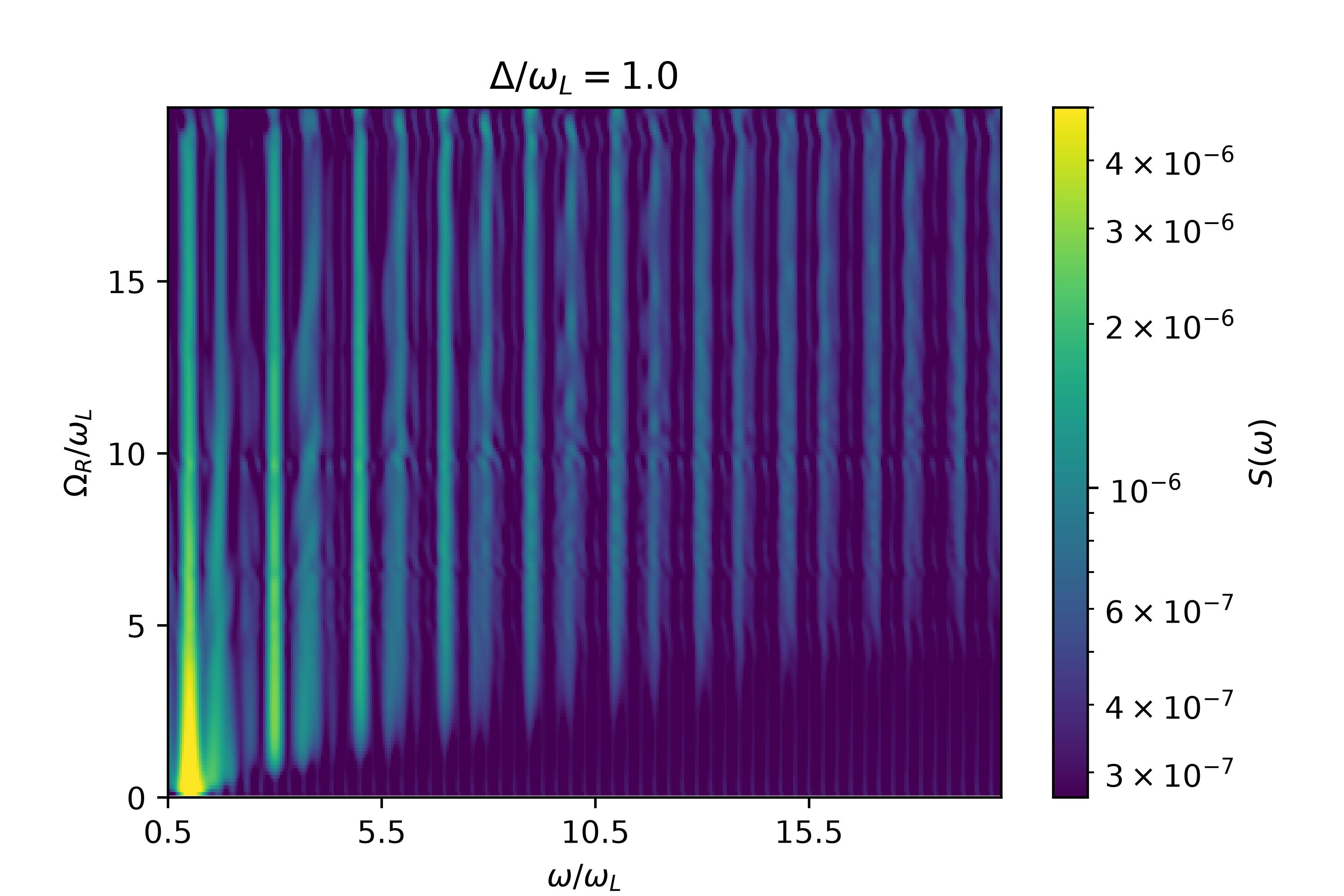}
\end{minipage}
\hfill
\begin{minipage}{0.48\textwidth}
    \centering
    \includegraphics[width=\linewidth]{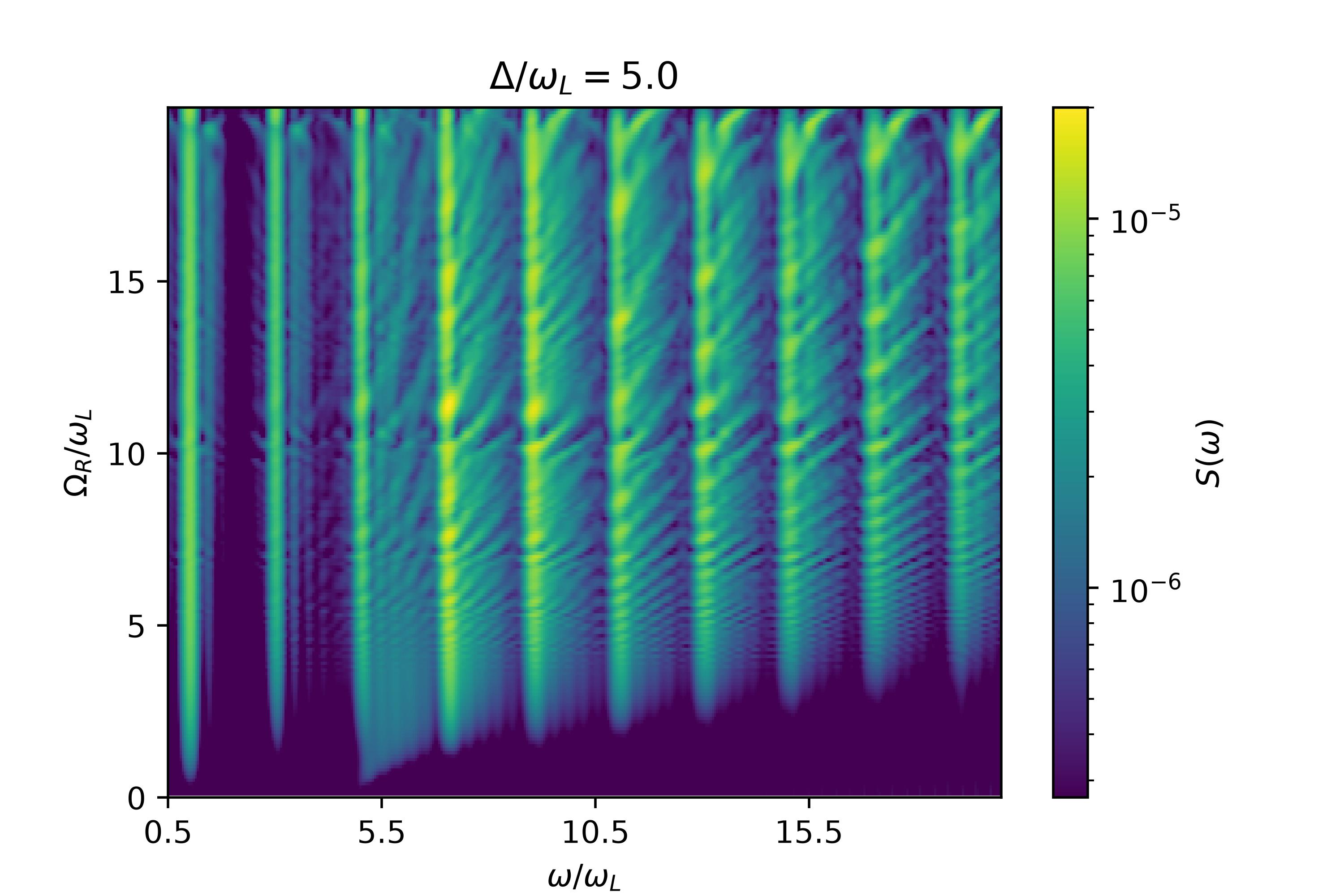}
\end{minipage}

\caption{Emission specturm $S(\omega)$ of a two-level system in and out of resonance (left and right panels, respectively) driven by coherent and squeezed vacuum (top and bottom panels respectively) for different Rabi intensities $\Omega_R$. The bottom panels are analog to those of~\cite{Tritschler2003}.}
\label{fig: cutoff-dependence}
\end{figure*}

We compare the HHG emission spectrum directly, the quantum optical way, i.e. by computing the quantized photon spectral occupation~\cite{Loudon2000, Scully1997} (not by calculating electronic dipoles~\cite{Gorlach2020, Stammer2024, Liu2026}),

\begin{align}
    \frac{dS(\omega_\eta)}{d\omega} = \frac{\omega_\eta \braket{ \hat{a}^\dagger_\eta \hat{a}_\eta }}{\Delta\omega},
    \label{eq: emission-int}
\end{align}

\noindent with $\eta = \omega/\omega_L$. For both a BSV and coherent driving~\cite{Tritschler2003, Hazanov2025}, we evaluate the emission spectrum for difference values of the Rabi frequency $\Omega_R$. The results are presented in Fig.~\ref{fig: cutoff-dependence}. The top panels show the emission spectra for coherent driving in the resonant (left) and nonresonant (right) regimes, while the bottom panels show the corresponding spectra for BSV driving with otherwise identical paramaters. In both case, the expectation values are obtained directly from the quantum state of the photon continuum given by Eq.~\eqref{eq: emission-int}, without relying on an additional reconstruction of the emitted field. One prominent feature of the coherent-driving spectra is the oscillatory structure associated with the Floquet eigenenergies in the resonant regime (left panels), which is absent under BSV driving~\cite{Tritschler2003, Hazanov2025}. In addition, the cutoff energy exhibits a stronger dependence on the Rabi frequency in the BSV case. This enhanced cutoff dependence is consistent with previous studies in atoms and solids~\cite{Gorlach2023, Gothelf2025}, where such cutoff dependence was identified through an analysis of the Husimi distribution of the BSV. We suspect that the blurred-out spectral features in the BSV case are a result of its broad photon statistics. This essentially applies something akin to an effective intensity and phase averaging in every spectral point, masking finer details. \\

%% Figure of the correlation in continuum
%% g^(2) figure
%% R parameter figure

\noindent Next, we compute the second- and third-order correlation spectra given by

\begin{align}
    \mathcal{C}^{(2)}(\omega_1,\omega_2) = \frac{d^2 \mathrm{Corr}(\omega_1, \omega_2)}{d\omega_1 d\omega_2} \approx \frac{\braket{ \hat{a}^\dagger_{\eta_1} \hat{a}^\dagger_{\eta_2} \hat{a}_{\eta_1} \hat{a}_{\eta_2} }}{\Delta\omega^2},
\end{align}

\begin{align}
    \mathcal{C}^{(3)}(\omega_1,\omega_2, \omega_3) = \frac{d^2 \mathrm{Corr}(\omega_1, \omega_2, \omega_3)}{d\omega_1 d\omega_2 d\omega_3} \approx \frac{\braket{ \hat{a}^\dagger_{\eta_1} \hat{a}^\dagger_{\eta_2} \hat{a}^\dagger_{\eta_3} \hat{a}_{\eta_1} \hat{a}_{\eta_2} \hat{a}_{\eta_3} }}{\Delta\omega^3},
\end{align}

\begin{figure*}[ht]
    \centering
    \includegraphics[width=0.5\linewidth]{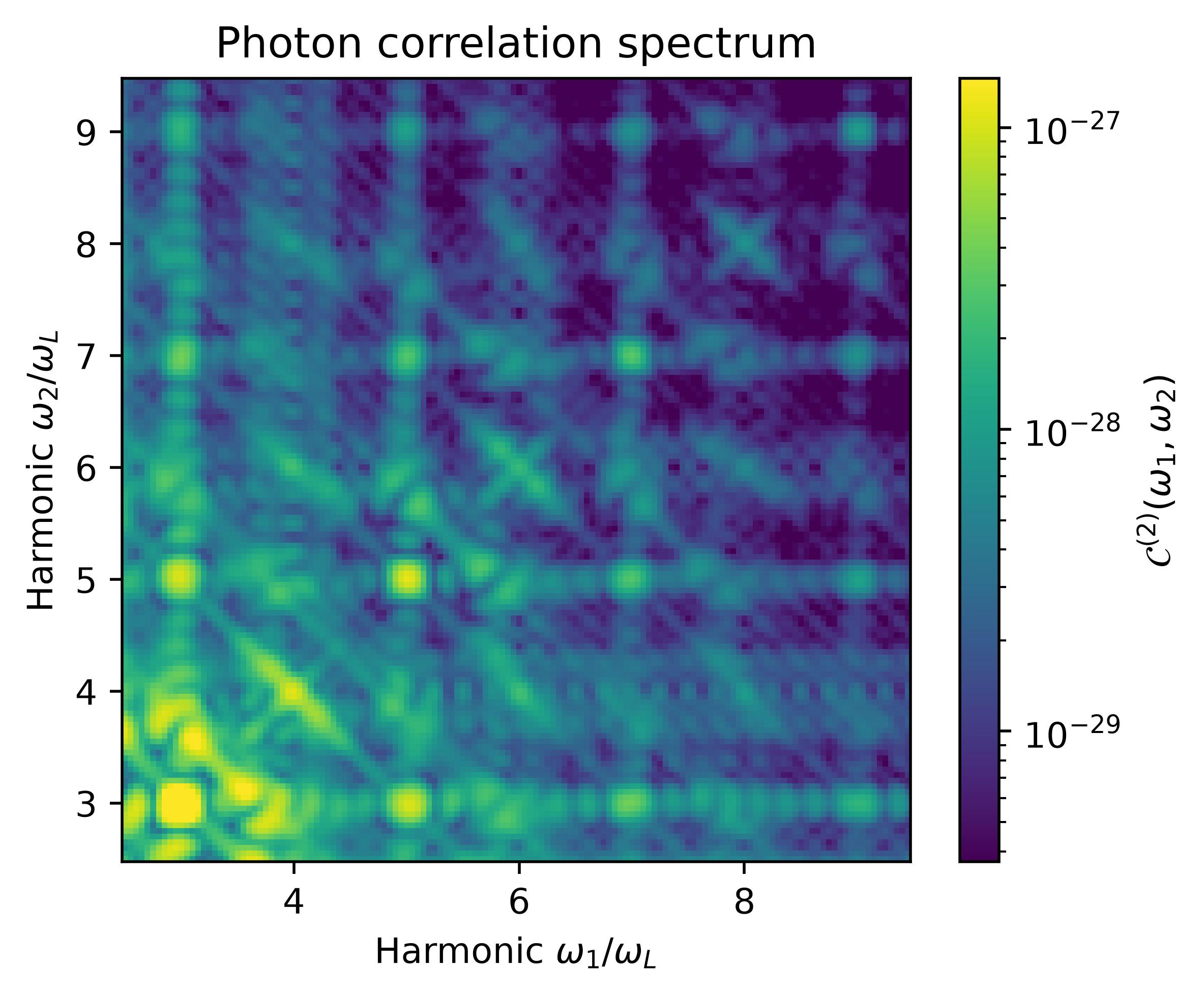}
    \caption{Second-order photon correlation spectrum for the continuum of emission modes $\mathcal{C}^{(2)}(\omega, \omega^\prime)$.}
    \label{fig: second-correlation-spectrum}
\end{figure*}

\begin{figure*}[ht]
    \centering
    \includegraphics[width=\linewidth]{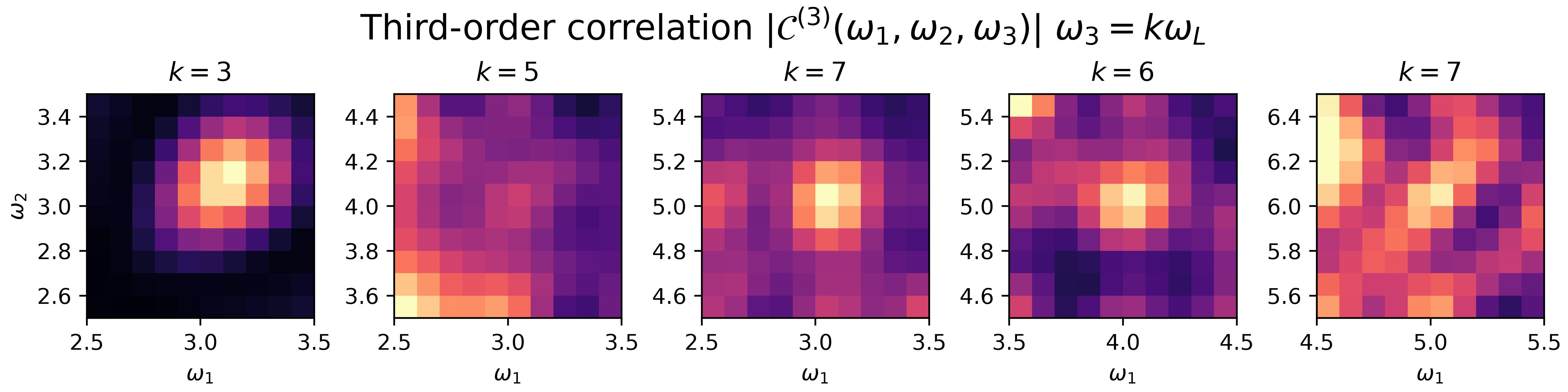}
    \caption{Exemplary panels of the third-order photon correlation spectrum of the continuum of emission modes $\mathcal{C}^{(3)}(\omega_1, \omega_2, \omega_3)$. The third frequency is fixed to the $k$th harmonic frequency, $\omega_3 = k\omega_L$, while each panel shows the correlation spectrum as a function of the first and second frequencies, $\omega_1$ and $\omega_2$, in the vicinity of the $i$th and the $j$th harmonics.}
    \label{fig: correlation-third-examples}
\end{figure*}

\noindent with $\eta_k = \omega_k/\omega_L$. Figs.~\ref{fig: second-correlation-spectrum} and~\ref{fig: correlation-third-examples} show second- and third-order correlations spectra (for every simulation three-cycle pulses were used and the Rabi intensity was $\Omega_R/\omega_L = 2$). Such correlation measurements are emerging as promising tools for quantum metrology~\cite{Theidel2024, Theidel2025} and quantum spectroscopy based on HHG~\cite{Lemieux2025, Lyu2026}, yet their theoretical characterization in this regime remains largely unexplored. As in the intensity spectra of Fig.\ref{fig: cutoff-dependence}, both odd and even harmonics are present owing to the broken axial symmetry of the TLS. Their correlation signatures, however, are strikingly different. In Fig.~\ref{fig: second-correlation-spectrum}, odd harmonics exhibit predominantly ring-like resonance structures, whereas even harmonics develop characteristic star-like patterns. The third-order correlations in Fig.~\ref{fig: correlation-third-examples} reveal an even richer landscape, with pronounced symmetry-broken structures and no simple universal pattern across harmonic orders. A complete map of the relevant third-order mode correlations is provided in the Supplementary Information, offering concrete predictions for future experiments. These results suggest that higher-order photon correlations encode information inaccessible to intensity or second-order measurements, potentially providing a more sensitive probe of light–matter dynamics and symmetry in quantum HHG.

\subsection{Wigner negativities and back-action}

\begin{figure*}[t]
\centering

\begin{minipage}{0.48\textwidth}
    \centering
    \includegraphics[width=0.9\linewidth]{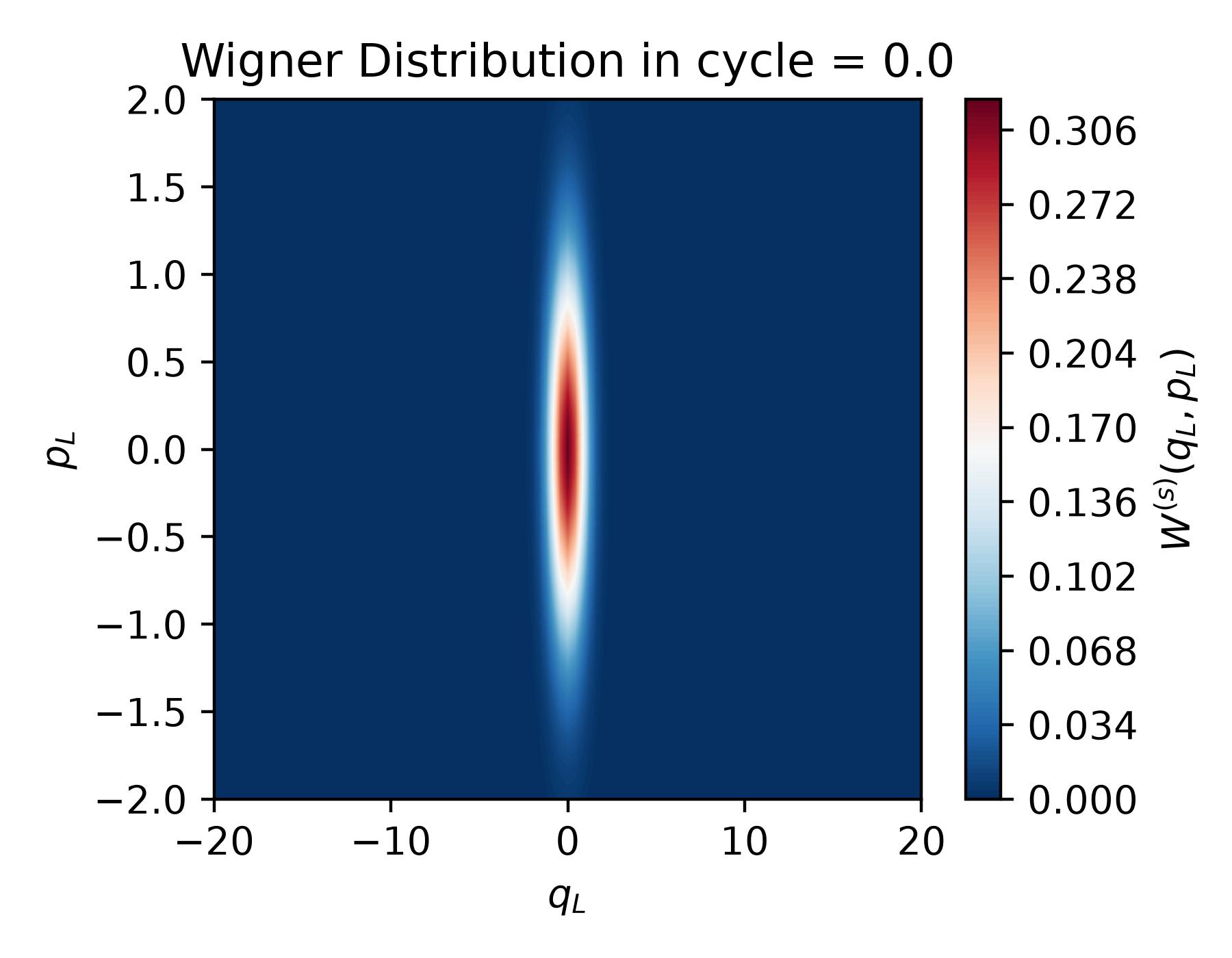}
\end{minipage}
\hfill
\begin{minipage}{0.48\textwidth}
    \centering
    \includegraphics[width=0.9\linewidth]{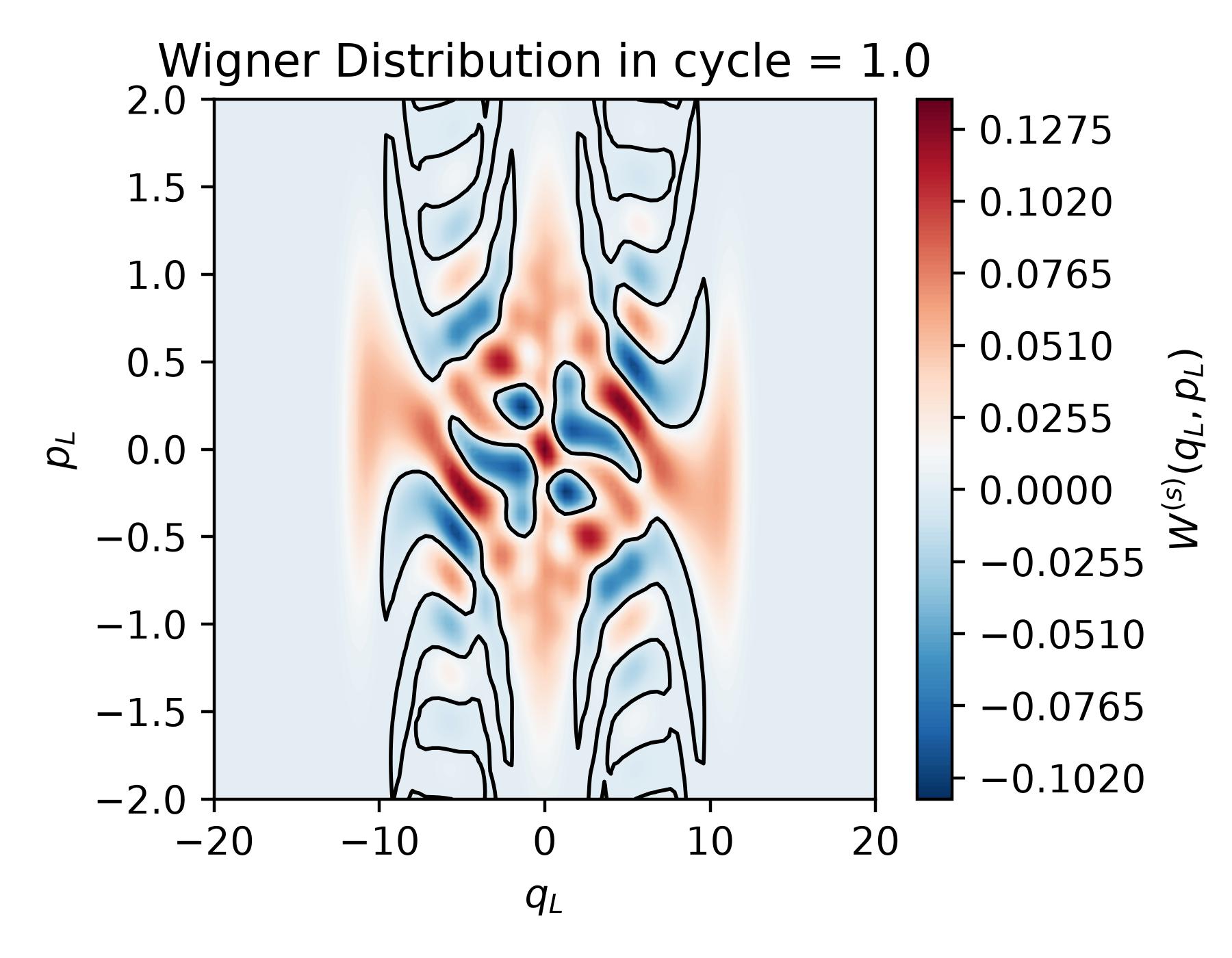}
\end{minipage}

\vspace{0.1cm}

\begin{minipage}{0.48\textwidth}
    \centering
    \includegraphics[width=0.9\linewidth]{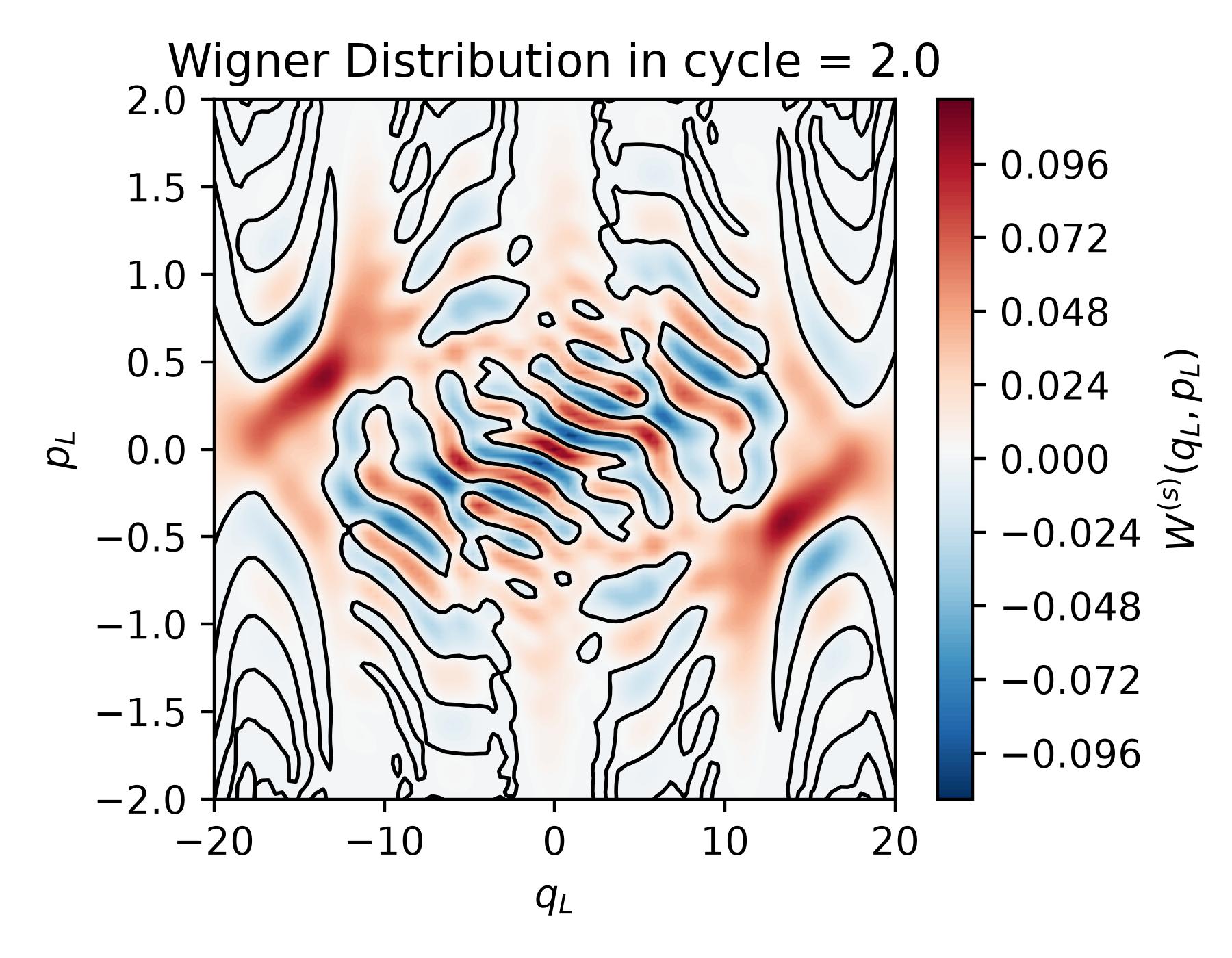}
\end{minipage}
\hfill
\begin{minipage}{0.48\textwidth}
    \centering
    \includegraphics[width=0.9\linewidth]{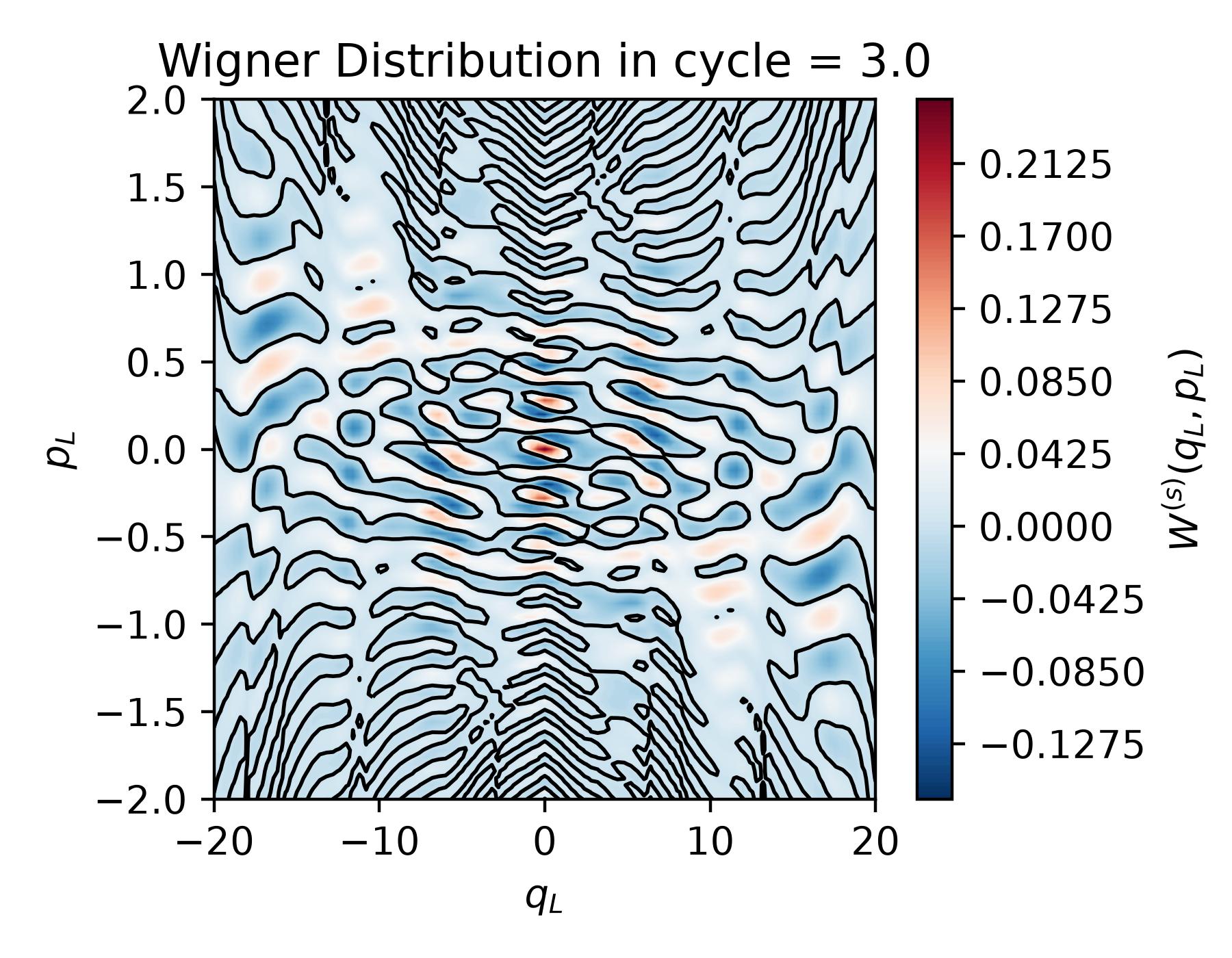}
\end{minipage}
\caption{Wigner distribution in the squeezed picture $W^{(s)}({\bf x}_L;t)$ for the driving mode over time. Snapshots of at $t = 0,~\tau_L,~2\tau_L,~3\tau_L$ are shown. The black lines represent transitions from positive to negatives values of the Wigner distribution.}
\label{fig: wigner-dist}
\end{figure*}

\begin{figure*}[t]
\centering

\begin{minipage}{0.48\textwidth}
    \centering
    \includegraphics[width=0.9\linewidth]{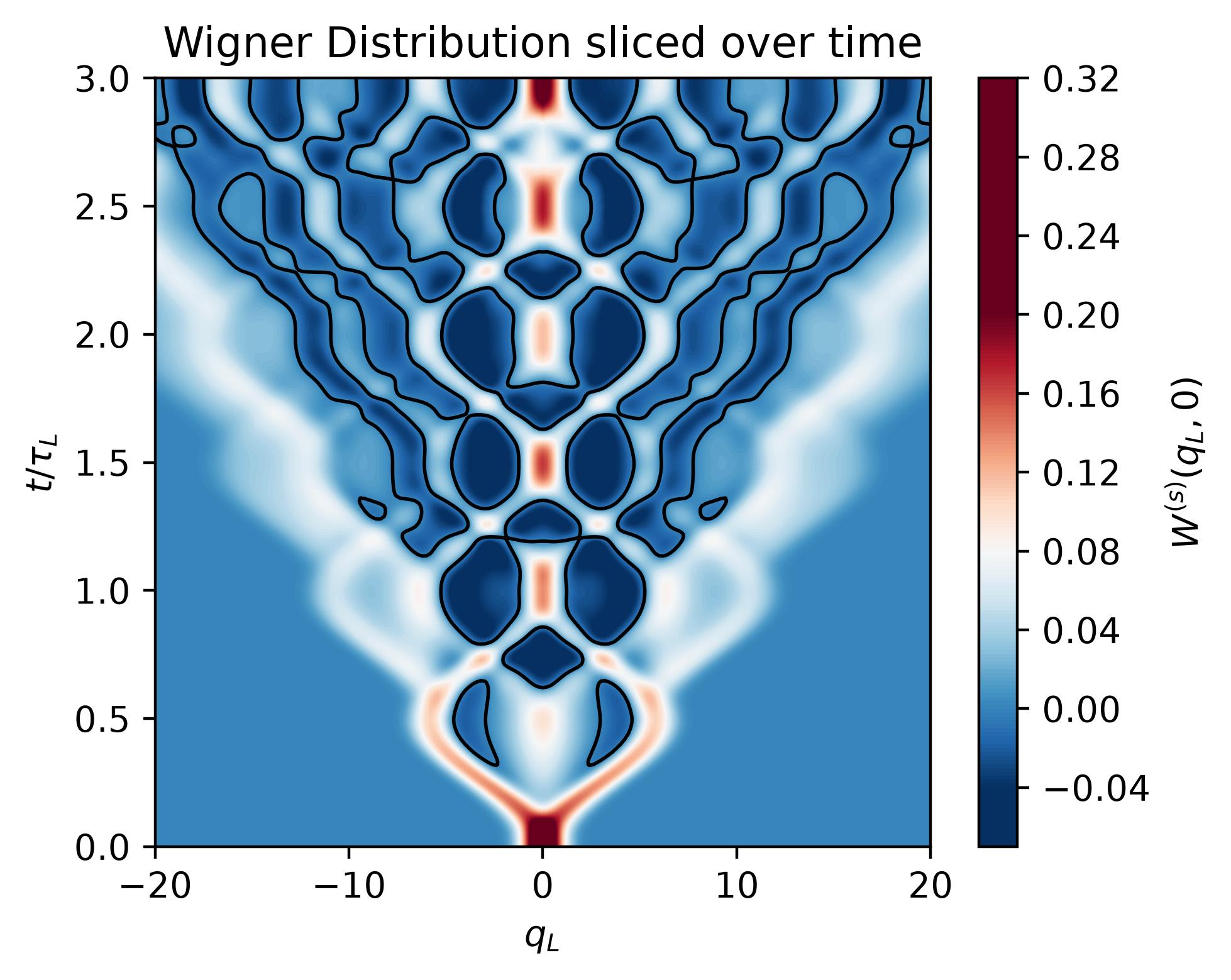}
\end{minipage}
\hfill
\begin{minipage}{0.48\textwidth}
    \centering
    \includegraphics[width=0.9\linewidth]{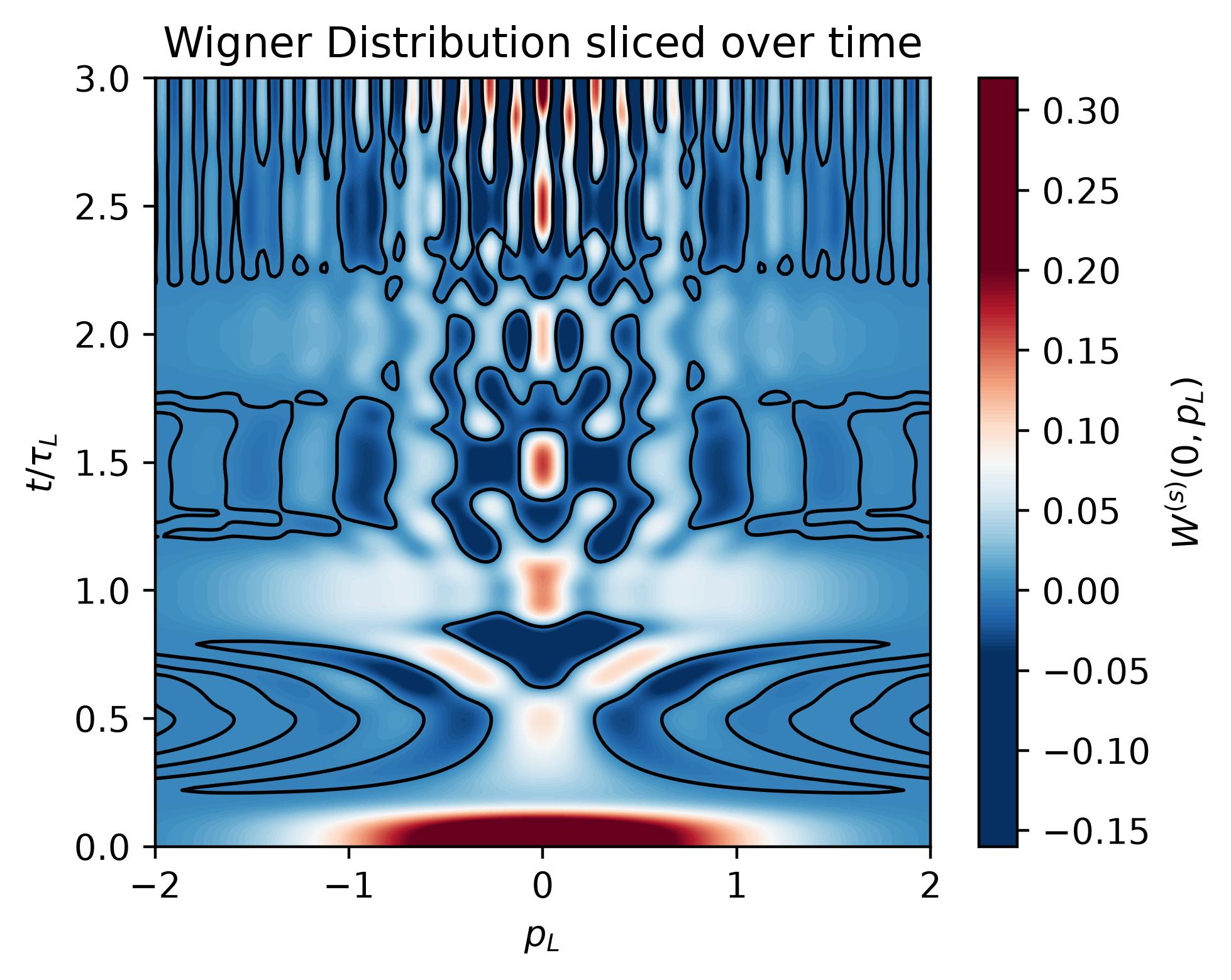}
\end{minipage}
\caption{Time evolution of the position (left) and momentum (right) slices of the Wigner distribution in the squeezed picture $W^{(s)}({\bf x}_L;t)$ for the driving mode over time, $W^{(s)}(q_L, 0;t)$ and $W^{(s)}(0, p_L;t)$, respectively.}
\label{fig: wigner-slice}
\end{figure*}

Figure~\ref{fig: wigner-dist} presents the quantum state of the driving mode through the Wigner distribution in the squeezed-interaction picture

\begin{align}
    W^{(s)}(q_L, p_L;t) \equiv \frac{1}{\pi} \int_{-\infty}^{\infty} \bra{q_L-y_L} \hat{\rho}^{(s)}_L(t) \ket{q_L+y_L} e^{2ip_Ly_L} dy_L = \sum_{n,m} \bra{n}\hat{\rho}^{(s)}_L(t) \ket{m} w_{nm}(q_L, p_L),
\end{align}

\noindent where $(q_L, p_L)$ denotes the Wigner phase-space coordinate of the driver in the squeezed-interaction picture, $\hat{\rho}^{(s)}_L(t) = \mathrm{Tr}_{\mathrm{TLS}, \mathrm{cont}} \left\lbrace \ket{\Psi^{(s)}(t)} \bra{\Psi^{(s)}(t)} \right\rbrace$ is the traced density matrix of the driver (tracing out the degrees of freedom of the electronc and the continuum of modes), $n$ and $m$ are Fock state indices in the squeezed-interaction picture, and $w_{nm}(q_L, p_L)$ is the Wigner distribution for the Fock matrix element $\ket{n}\bra{m}$. Additionally, Fig.~\ref{fig: wigner-slice} presents the time evolution of the Wigner distribution in slices of position and momentum, $W^{(s)}(q_L,0,t)$ and $W^{(s)}(0,p_L,t)$, respectively (see SI for full movie with attosecond temporal resolution). Notably, we observe the emergence of Wigner-negative regions as the squeezed intense laser drives the TLS. By contrast, a semiclassical treatment of the squeezed driving field preserves a positive Wigner distribution by construction and therefore cannot capture these features. The emergence of Wigner negativity thus provides a direct signature of genuinely quantum light–matter dynamics, beyond descriptions based solely on semiclassical phase-space statistics, and offers an experimentally accessible probe of quantum back-action. We note that, although the distribution deviates significantly from bare BSV as time progresses, the driving mode remains strongly squeezed throughout the evolution. This apparent discrepancy arises because the phase-space coordinates are defined in the squeezed picture and therefore already incorporate the squeezing transformation. \\ 

\begin{figure}
    \includegraphics[width=0.75\linewidth]{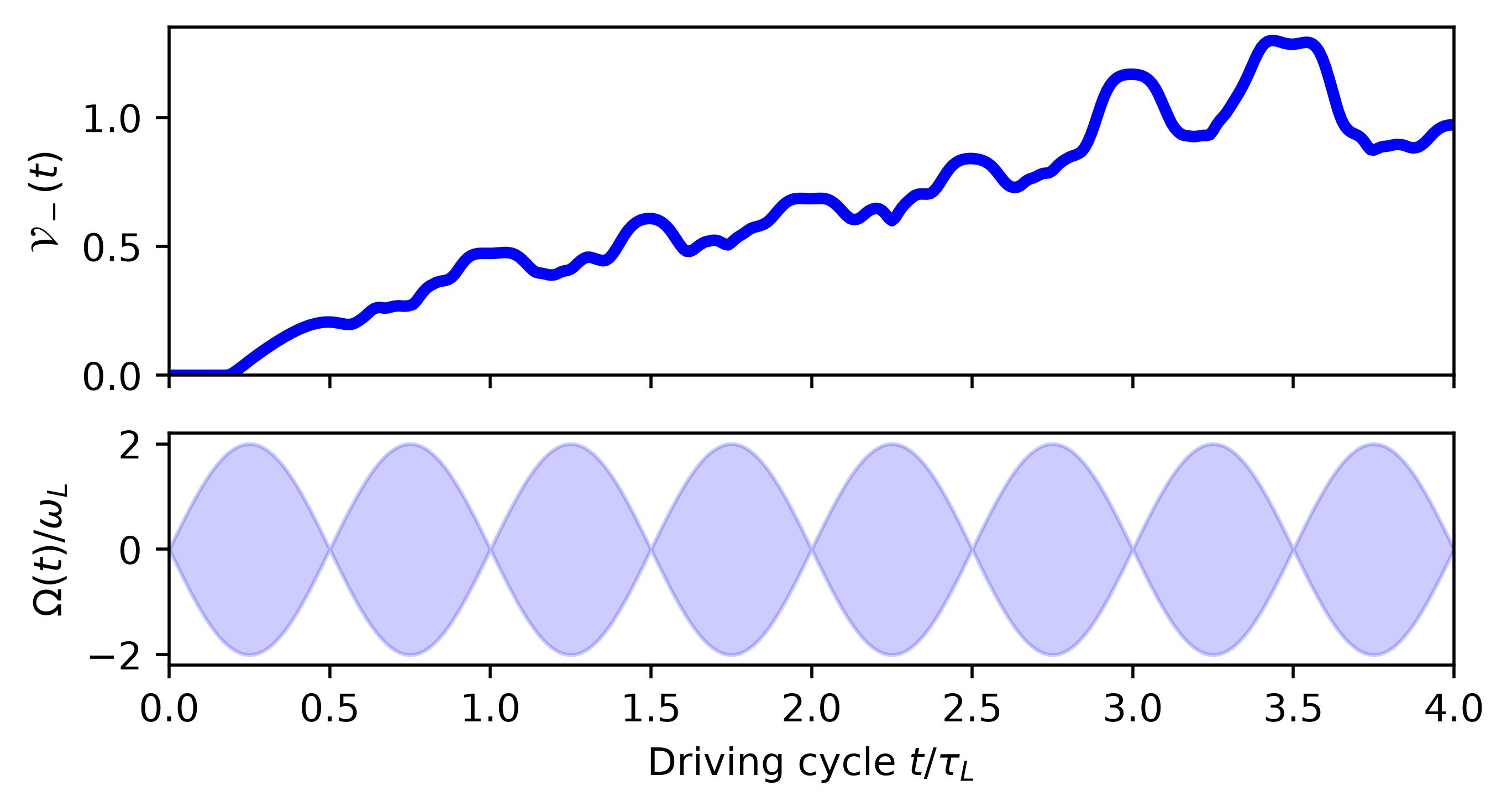}
    \caption{Wigner-negative volume $\mathcal{V}_-(t)$ over time (top panel) and time-dependent Rabi amplitude of the driver, $\Omega(t) = \Omega_R \left| \sin{(\omega_L t)} \right| $.}
    \label{fig: negativity}
\end{figure}

\noindent To quantify the degree of nonclassicality generated during the evolution, especially its temporal dynamics, we employ the negativity volume measure~\cite{Arkhipov2018}, which is invariant under the squeezeing transformation (see SI for details),

\begin{align}
    \mathcal{V}_{-}(t) = -\int_{\mathcal{D}_-} W(q_L,p_L;t) dq_L dp_L,
\end{align}

\noindent where $\mathcal{D}_-$ denotes the region of phase space in which the Wigner distribution is negative. Figure~\ref{fig: negativity} shows the resulting Wigner negativity volume as a function of time (top panel), as well as the time-dependent Rabi frequency defined as the instantaneous electric field variance $\Omega(t) = \mu \sqrt{\bra{\xi}\hat{E}_{\mathrm{L}}^2(t) \ket{\xi}} = \Omega_R \left| \sin{(\omega_L t)} \right|$. The negativity volume exhibits pronounced local maxima at half-cycle intervals, which coincide with the minimal field variance of the driver. This correlation indicates that the strongest nonclassical features of the driving mode are generated at the end of the strong-interaction periods, phase shifted from the peak electric field (and following the vector potential). \\

\noindent Notably, the Wigner negativity develops on attosecond timescales and exhibits a strongly nonlinear temporal evolution, highlighting the nonperturbative character of the quantum back-action. As shown in Fig.~\ref{fig: negativity}, this dynamics contains contributions across multiple harmonic orders, reflecting the nonlinear response of the driven system. The nonclassicality also accumulates with interaction time: the full temporal evolution in the Supplementary Information shows that each additional laser half-cycle generates a new negative lobe in the Wigner distribution of the driving mode. This suggests that the emergence of quantum features is governed not only by the strength and nonlinearity of the light–matter interaction, but also by its spatiotemporal extent. Resolving this ultrafast buildup of nonclassicality therefore calls for quantum-optical probes with attosecond temporal resolution.

%% Figure occupation (show also MTEF)

%%%%%%%%%%%%%%%%%%%%%%%%%%%%%%%%%%%%%%%%%%%%%%%%%%%%%%%%%%%%%%%%%%
%                        Conclusions and outlook                            %
%%%%%%%%%%%%%%%%%%%%%%%%%%%%%%%%%%%%%%%%%%%%%%%%%%%%%%%%%%%%%%%%%%

\section{Conclusions and outlook}

In this work, we fully propagate the coupled light–matter wavefunction of a two-level system driven by bright squeezed vacuum and interacting with a continuum of quantized emission modes. A squeezed-frame transformation greatly accelerates convergence of the driving-field Hilbert space, enabling access to the large-photon-number regime. We compute the resulting HHG spectrum and benchmark it against coherent driving. Bright-squeezed-vacuum excitation produces a markedly stronger dependence of the HHG cutoff on Rabi intensity, while other spectral features are broadened or washed out by the intrinsic phase and intensity fluctuations of the squeezed field. \\

\noindent Furthermore, our method enables the direct calculation of second- and third-order photon-field correlations, revealing distinct resonances in the vicinity of the emitted harmonics. We also reconstruct the Wigner distribution of the driver and show that back-action causes: (i) negative-valued regions to emerge naturally from the self-consistent coupling between light and matter; (ii) The negativity volume reaches its maximum at half-cycles, coinciding with the vector potential maxima; (iii) The negativity evolves on attosecond timescales, and truly quantum features generally increase substantially over interaction time; (iv) The effect is highly nonlinear and the back-action develops harmonics of the drive.\\

\noindent This work provides the first fully self-consistent propagation of a matter system coupled simultaneously to a quantized driver and fully quantized spectrum of emission modes. The framework can potentially be extended to more complex matter systems and to other nonclassical quantum drivers. Moreover, the analysis of the backaction of matter on the driver may provide a pathway toward the development of experimental techniques for characterizing squeezed pulses. While we employed here TLS as a more phenomenological model that enables to tractably propagate the electron-photon system, we expect that it might find direct usage in analogous nearly-TLS such as diamond NV centers, excitons, systems with spatially-localized flat bands, etc.

%%%%%%%%%%%%%%%%%%%%%%%%%%%%%%%%%%%%%%%%%%%%%%%%%%%%%%%%%%%%%%%%%%
%                        Acknowledgments                            %
%%%%%%%%%%%%%%%%%%%%%%%%%%%%%%%%%%%%%%%%%%%%%%%%%%%%%%%%%%%%%%%%%%

\section*{Acknowledgements}

This work was funded by the European Union under the ERC Synergy Grant UnMySt (HEU GA No. 101167294). Views and opinions expressed are however those of the author(s) only and do not necessarily reflect those of the European Union or the European Research Council. Neither the European Union nor the European Research Council can be held responsible for them. This work was also supported by the Cluster of Excellence ‘Advanced Imaging of Matter’ (AIM), Grupos Consolidados (IT1453-22) and Deutsche Forschungsgemeinschaft (DFG) - SFB-925 - project 170620586. The Flatiron Institute is a division of the Simons Foundation. We acknowledge support from the Max Planck-New York City Center for Non-Equilibrium Quantum Phenomena. S.d.l.P. acknowledges support from International Max Planck Research School. M.F.C. acknowledges support by the Quantum Science and Technology-National Science and Technology Major Project (Grant No. 2025ZD0301000),  the National Key Research and Development Program of China (Grant No. 2023YFA1407100), the Guangdong Province Science and Technology Major Project (Future functional materials under extreme conditions - 2021B0301030005) and the National Natural Science Foundation of China (Grant No. 12574092). O.N. acknowledges support of the Young Faculty Award from the National Quantum Science and Technology program of Israel's Council of Higher Education Planning and Budgeting Committee and support from The Technion Helen Diller Quantum Center and RBNI through the Nevet grant. The authors are grateful to M.F.C. for hosting a research visit at the Guangdong Technion-Israel Institute of Technology during March 2026, where this project was conceived.

%%%%%%%%%%%%%%%%%%%%%%%%%%%%%%%%%%%%%%%%%%%%%%%%%%%%%%%%%%%%%%%%%%
%                        Bibliography                            %
%%%%%%%%%%%%%%%%%%%%%%%%%%%%%%%%%%%%%%%%%%%%%%%%%%%%%%%%%%%%%%%%%%

%

\newpage
\onecolumngrid

\appendix
\renewcommand{\thefigure}{S\arabic{figure}}
\setcounter{figure}{0}

\section{Light-matter coupling for a one-dimensional waveguide}

\noindent For a one-dimensional model of light and matter there is an ambiguity in the way the problem is phrased. Given a waveguide of sectional area $A_{\mathrm{eff}}$, the light-matter coupling would be given by

\begin{align}
    \hat{H}_{\mathrm{cont}}(t) = \hat{\sigma}_x \sum_{k,s} g_{k,s} \left[ \hat{a}_{k,s} e^{-i \omega_k t} + \hat{a}^\dagger_{k,s} e^{i \omega_k t}  \right]
\end{align}

\begin{align}
    g_{k,s} = \pmb{\mu}\cdot \pmb{e}_{k,s} \sqrt{\frac{\hbar \omega_k}{2 A_{\mathrm{eff}} L \epsilon_0 }} = \pmb{\mu}\cdot \pmb{e}_{k,s} \sqrt{\frac{2\pi \omega_k}{A_{\mathrm{eff}} L}}
\end{align}

\noindent where $L$ is the length of the waveguide (which we orient to the z-direction) which we will take to be infinite at the end of our calculation, $\pmb{\mu} = \mu \pmb{e}_x$ is the dipole moment of matter and is fixed in the x-direction as it is the orientation of the driver. Also, since we only consider the modes propagation in the longitudinal direction of the waveguide, we have that $\omega_k = \left| k \right| c = \left| k \right|/\alpha$, where $\alpha$ is the fine structure constant and $k$ is an integer. From the two-polarizations of light, oriented in x and y, only those that match the x-direction are coupled to the matter system since $\pmb{\mu}  \cdot \pmb{e}_y =0$, thus $g_{k,y} = 0$ and we only consider those polarizations oriented in the x-direction, which have $g_{k,x} = \mu \sqrt{2\pi \omega_k/A_{\mathrm{eff}} L}$. The mode volume element is therefore given by $v_\mathrm{m} = 2\pi/L$, the frequency spacing is $\Delta\omega = v_\mathrm{m}c =  v_\mathrm{m}/\alpha$, and we can integrate over the wavenumber in continuum for $L$ large enough to obtain an effective coupling to frequency modes, $g_\eta$, each labeled by $\eta = \omega/\omega_L$:

\begin{align}
    \hat{H}_{\mathrm{cont}}(t) = \hat{\sigma}_x \sum_{\eta\in \mathcal{D}} g_\eta \left[ \hat{a}_\eta e^{-i\omega_\eta t} + \hat{a}_\eta^\dagger e^{i\omega_\eta t}  \right],
    \label{1D-coupling}
\end{align}

\begin{align}
    g_\eta = \sqrt{ \sum_{k} g_{k,x}^2 \delta\left( \omega_\eta -\frac{\left| k \right|}{\alpha} \right) \Delta\omega }  \approx \mu \sqrt{ \frac{2\pi}{ L} \int_{-\infty}^\infty \frac{  \omega}{A_{\mathrm{eff}}} \delta\left( \omega_\eta - \left|\omega\right| \right) d\omega } = 2\mu \sqrt{\frac{\alpha}{A_{\mathrm{eff}}}} \sqrt{\Delta\omega} \sqrt{ \frac{\omega_\eta}{2}}.
\end{align}

\noindent Now we can take $\lambda = g \sqrt{\Delta \omega}$ as the light-matter coupling, and $g = 2\mu \sqrt{\alpha / A_{\mathrm{eff}}}$ is the spectral light matter coupling strength. The factor $g$ contains the effective coupling of matter with the energy mode.

\begin{align}
    \hat{H}_{\mathrm{cont}}(t) = \hat{\sigma}_x \sum_{\eta\in \mathcal{D}} \lambda \sqrt{\frac{\omega_\eta}{2}} \left[ \hat{a}_\eta e^{-i\omega_\eta t} + \hat{a}_\eta^\dagger e^{i\omega_\eta t}  \right]
    \label{1D-coupling}
\end{align}

\noindent where  Equation~\eqref{1D-coupling} is the continuum interaction Hamiltonian from Eq.~\eqref{eq: continuum-ham}. \\

\noindent To minimize computational cost, we restrict the spectral domain $\mathcal{D}$, that is, the set of quantized continuum modes in Eq.~\eqref{eq: continuum-ham}, to the smallest mode space required for convergence in each calculation. Consequently, to compute spectra such as those in Figs.~\ref{fig: cutoff-dependence},~\ref{fig: correlation-spectrum}, and~\ref{fig: correlation-third-examples}, we isolate only the relevant harmonic windows. For instance, to calculate the correlation bewteen the $n$th and $m$th harmonics we restrict the spectral domain to $\mathcal{D} = \left\lbrace \eta\ \mid\, \eta\in \left( n-0.5,~n+0.5 \right) \cup \left( m-0.5,~m+0.5  \right) \right\rbrace $. This makes the calculations faster and avoids the need for renormalization of the TLS gap energy $\Delta$. Including a larger range of continuum modes or enhancing the light-matter coupling by modifying the mode structure would require a more careful treatment of gap renormalization.

\section{Rabi frequency for coherent and squeezed states}

\noindent In order to compare coherent and squeezed driving using a consistent definition of Rabi frequency we will associate it directly with the laser power for both states, that is $\Omega_R \sim \sqrt{I_{\mathrm{laser}}}$. Hence, we define the Rabi frequency of a quantum laser with period $T$ as

\begin{align}
    \Omega_R \equiv \sqrt{2}\mu E_{\mathrm{rms}} = \mu \sqrt{\frac{2}{T} \int \braket{:\hat{E}_L^2(t):} dt}
    \label{eq: rabi}
\end{align}

\noindent with $\hat{E}_L(t) = \mathcal{E}_0 \left[ \hat{a}_L e^{-i\omega_Lt} + \hat{a}_L^\dagger e^{i\omega_L t} \right]$ the interaction picture quantum electric field operator, and $\mathcal{E}_0 = \sqrt{2\alpha \omega_L \Delta \omega / A_{\mathrm{eff}}} = \kappa/\mu $ is the electric field of a single driving-mode photon. The definition from Eq.~\eqref{eq: rabi} can be employed for any quantum state of light. Since the integral over time removes any time-dependent term, we end up with the time-independent terms containing $\hat{a}_L$ and $\hat{a}^\dagger_L$, such that

\begin{align}
    \frac{1}{T} \int:\hat{E}_L^2(t): dt = \kappa^2 :\left[ \hat{a}_L \hat{a}_L^\dagger + \hat{a}_L^\dagger \hat{a}_L  \right]: = 2 \mathcal{E}_0^2  \hat{a}_L^\dagger \hat{a}_L .
\end{align}

\begin{align}
    \Omega_R = 2 \kappa \sqrt{ \braket{\hat{a}_L^\dagger \hat{a}_L} }
\end{align}

\noindent For a non-interacting electromagnetic field, the photon number $\braket{\hat{a}_L^\dagger \hat{a}_L}$ is a constant of motion representing the energy of the field. Now, we can define the Rabi frequency for coherent, BSV and Fock drivings: $\ket{\alpha}_L$, $\ket{\xi}_L$, and $\ket{n}_L$, respectively; employing $\bra{\alpha}_L \hat{a}_L^\dagger \hat{a}_L \ket{\alpha}_L = \left| \alpha \right|^2$, $\bra{\xi}_L \hat{a}_L^\dagger \hat{a}_L \ket{\xi}_L = \sinh^2{\left| \xi \right|}$, $\bra{n}_L \hat{a}_L^\dagger \hat{a}_L \ket{n}_L = n$.

\begin{align}
    \Omega_R \left[ \ket{\alpha}_L \right] = 2\kappa \left| \alpha \right|,
\end{align}

\begin{align}
    \Omega_R \left[ \ket{\xi}_L \right] = 2 \kappa \sinh{\left| \xi \right|},
\end{align}

\begin{align}
    \Omega_R \left[ \ket{n}_L \right] = 2 \kappa\sqrt{n} .
\end{align}

\noindent Note that for $\left| \xi \right| \rightarrow \infty$, the BSV-associated Rabi frequency is $\Omega_R \left[ \ket{\xi}_L \right] \rightarrow \kappa \exp{(\left| \xi \right|)}$. Equation~\eqref{eq: driving-ham-bsv} represents the effective squeezed-picture driving Hamiltonian where the Rabi frequency just defined carries the information about the strength of the driving.

\section{Mode continuum spacing}

\begin{figure*}[ht]
    \centering
    \includegraphics[width=0.85\linewidth]{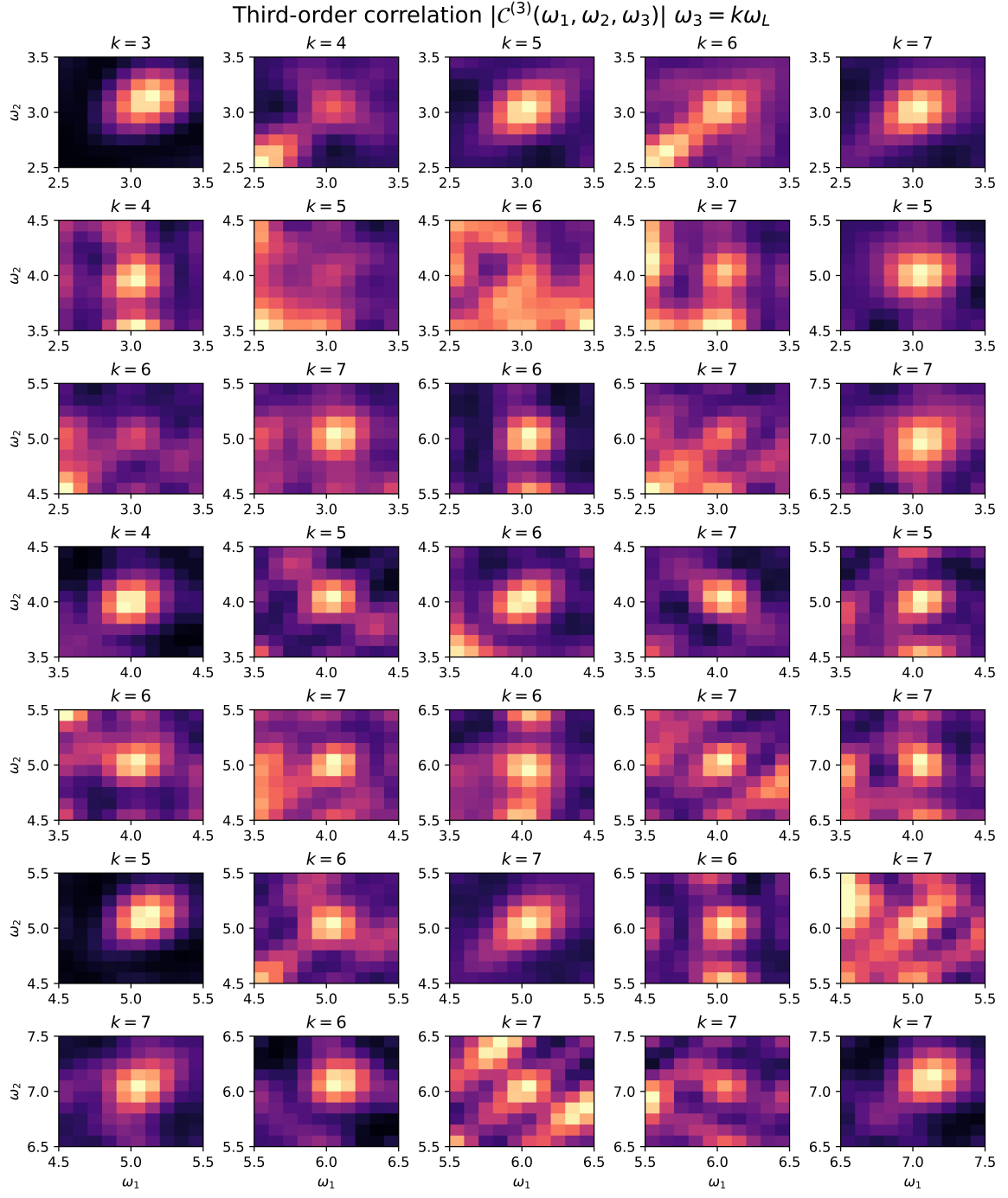}
    \caption{Third-order photon correlation spectrum for the continuum of emission modes $\mathcal{C}^{(3)}(\omega_1, \omega_2, \omega_3)$.}
    \label{fig: correlation-spectrum}
\end{figure*}

\noindent One of the most essential convergence parameters of this work is the mode spacing $\Delta \omega = \omega_L \Delta\eta$, which in Fig.~\ref{fig: correlation-spectrum} was picked to be $\Delta \eta = 0.05$. One can run the same calculation with $\Delta\eta = 0.1$
 and one gets the following result.

\begin{figure*}[ht]
    \centering
    \begin{minipage}{0.48\textwidth}
    \centering
    \includegraphics[width=\linewidth]{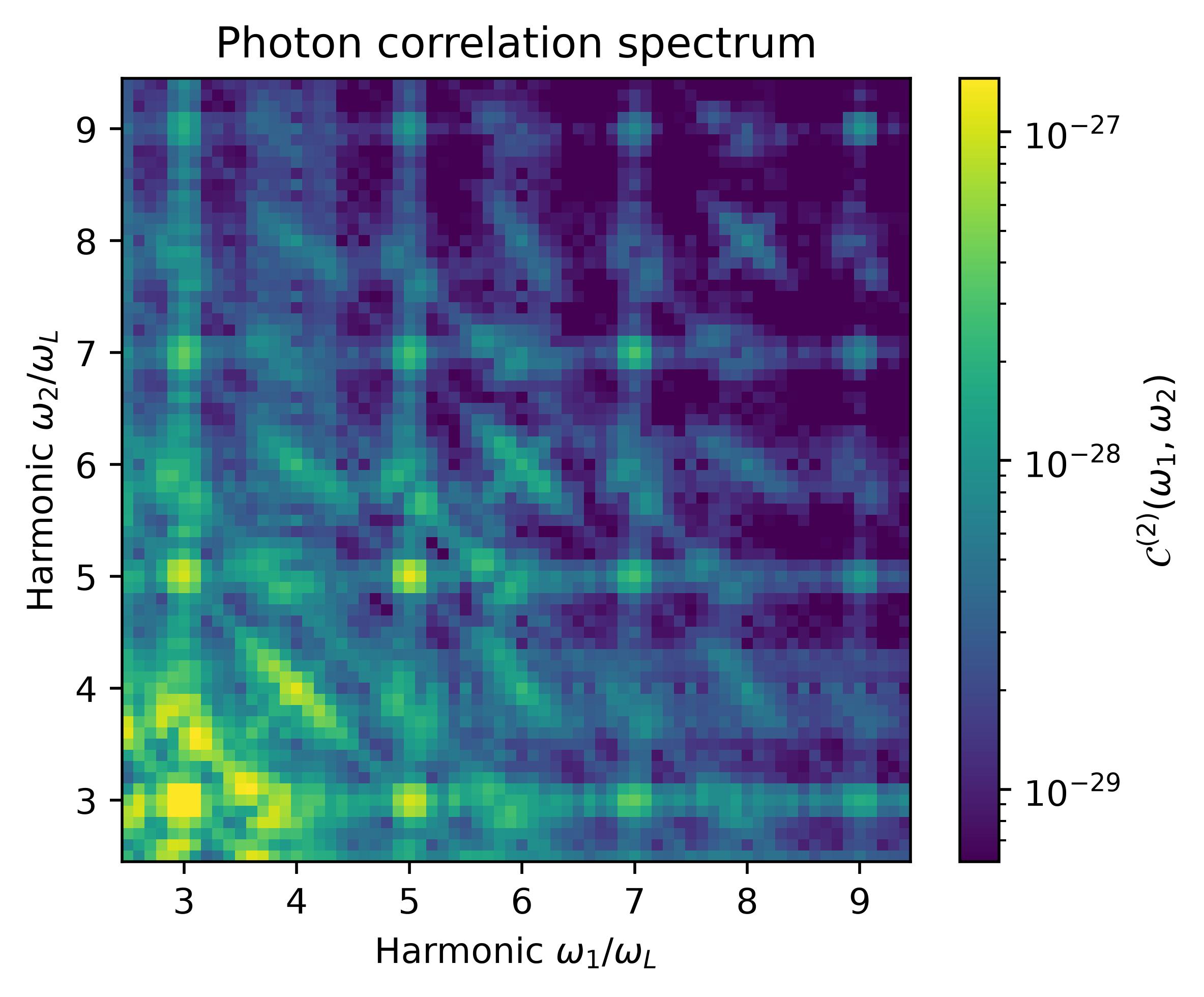}
\end{minipage}
\hfill
\begin{minipage}{0.48\textwidth}
    \centering
    \includegraphics[width=\linewidth]{figures/correlation-high-quality.jpg}
\end{minipage}
    
    \caption{Photon correlation spectrum for the continuum of emission modes $\mathcal{C}^{(2)}(\omega, \omega^\prime)$ with frequency spacing $\Delta \eta = 0.1$ (left) and $\Delta \eta = 0.05$ (right).}
    \label{fig: low-quality}
\end{figure*}

\section{Strong coherent driving Hamiltonian}

\noindent Eq.~\eqref{eq: driving-ham-bsv} is analogous to the displacement transformation of the driving Hamiltonian from Eq.~\eqref{eq: driving-ham}, 

\begin{align}
    \hat{H}^{(d)}_{\mathrm{driv}}(t) = \hat{D}^\dagger(\alpha) \hat{H}_{\mathrm{driv}}(t) \hat{D}(\alpha) = \mu \kappa \hat{\sigma}_x \left[ \hat{a}_L e^{-i\omega_Lt} + \hat{a}^\dagger_Le^{i\omega_L t} \right] + \Omega_R \hat{\sigma}_x \cos{(\omega_L t - \varphi)},
\end{align}

\noindent with $\Omega_R \left[ \ket{\alpha}_L \right] = 2 \kappa\left| \alpha \right|$ as the Rabi frequency and $\varphi$ the phase of the complex displacement $\alpha$. Note that in the limit of $\kappa \rightarrow 0$ and $\left| \alpha \right| \rightarrow \infty$ (infinite-photon limit) such that $\Omega_R$ remains constant, then the quantized field in the Hamiltonian (carrying the back-action information) can be neglected and only the external driving remains non-negligible:

\begin{align}
    \hat{H}^{(d)}_{\mathrm{driv}}(t) \rightarrow \Omega_R \hat{\sigma}_x \cos{(\omega_L t - \varphi)}.
    \label{eq: driving-ham-coh}
\end{align}

\noindent Eq.~\eqref{eq: driving-ham-coh} constrasts with Eq.~\eqref{eq: driving-ham-bsv} as in the latter case (BSV driving) the quantum back-action cannot be neglected, whereas in the former case (strong coherent driving) the external driving is completely inert to the emission of matter into the same driving mode.

\section{Semiclassical methods for BSV driving}

\noindent In semiclassical methods to run BSV calculations one can either employ the Wigner or the Husimi distribution and sample from it points in phase-space that act as initial conditions for different trajectories, $\left\lbrace \alpha^j_L(0) \right\rbrace$, where $j$ represents the trajectory label and $\alpha_L$ is the phase-space variable of the driving mode. The equations of motion for the light-matter system on each trajectory $\left\lbrace \alpha^j_L(t), \ket{\psi_j(t)} \right\rbrace$ are

\begin{align}
\begin{cases}
    i \ket{\dot{\psi}_j(t)} = \left[ \hat{H}_{\mathrm{TLS}} + \kappa \hat{\sigma}_x \Re \left\lbrace \alpha_L(t) \exp{(-i\omega_L t)} \right\rbrace \right]\ket{\psi_j(t)}, \\
    i\dot{\alpha}_L^j(t) = \kappa \bra{\psi_j(t)} \hat{\sigma}_x \ket{\psi_j(t)}\exp{(i\omega_L t)}/2,
\end{cases}
\end{align}

\noindent where $\kappa = \lambda \sqrt{\omega_L}$ combines all the prefactors for the light-matter coupling. Defining $\mathcal{E}(t) = \kappa \alpha(t)$ as the field phase-space variable and taking the limit $\kappa \rightarrow 0$ while preserving a distribution in $\mathcal{E}$ that is finite, we have

\begin{align}
\begin{cases}
    i \ket{\dot{\psi}_j(t)} = \left[ \hat{H}_{\mathrm{TLS}} + \hat{\sigma}_x \Re \left\lbrace \mathcal{E}_j(t) \exp{(-i\omega_L t)} \right\rbrace \right]\ket{\psi_j(t)}, \\
    i\dot{\mathcal{E}}_L^j(t) = \kappa^2 \bra{\psi_j(t)} \hat{\sigma}_x \ket{\psi_j(t)}\exp{(i\omega_L t)}/2 \rightarrow 0,
\end{cases}
\end{align}

\noindent which amounts to saying that the field can be treated as an external fixed amplitude and phase: $\dot{\mathcal{E}}_j(t) = 0 \rightarrow \mathcal{E}_j(t) = \mathcal{E}_j(0) = \left| \mathcal{E}_j \right| e^{i\varphi_j}$, such that the equations of motion simplify to

\begin{align}
    i \ket{\dot{\psi}_j(t)} = \left[ \hat{H}_{\mathrm{TLS}} + \hat{\sigma}_x \left| \mathcal{E}_j \right| \cos{(\omega_L t - \varphi_j)} \right]\ket{\psi_j(t)},
\end{align}
\begin{figure*}[ht]
    \centering
    \includegraphics[width=0.5\linewidth]{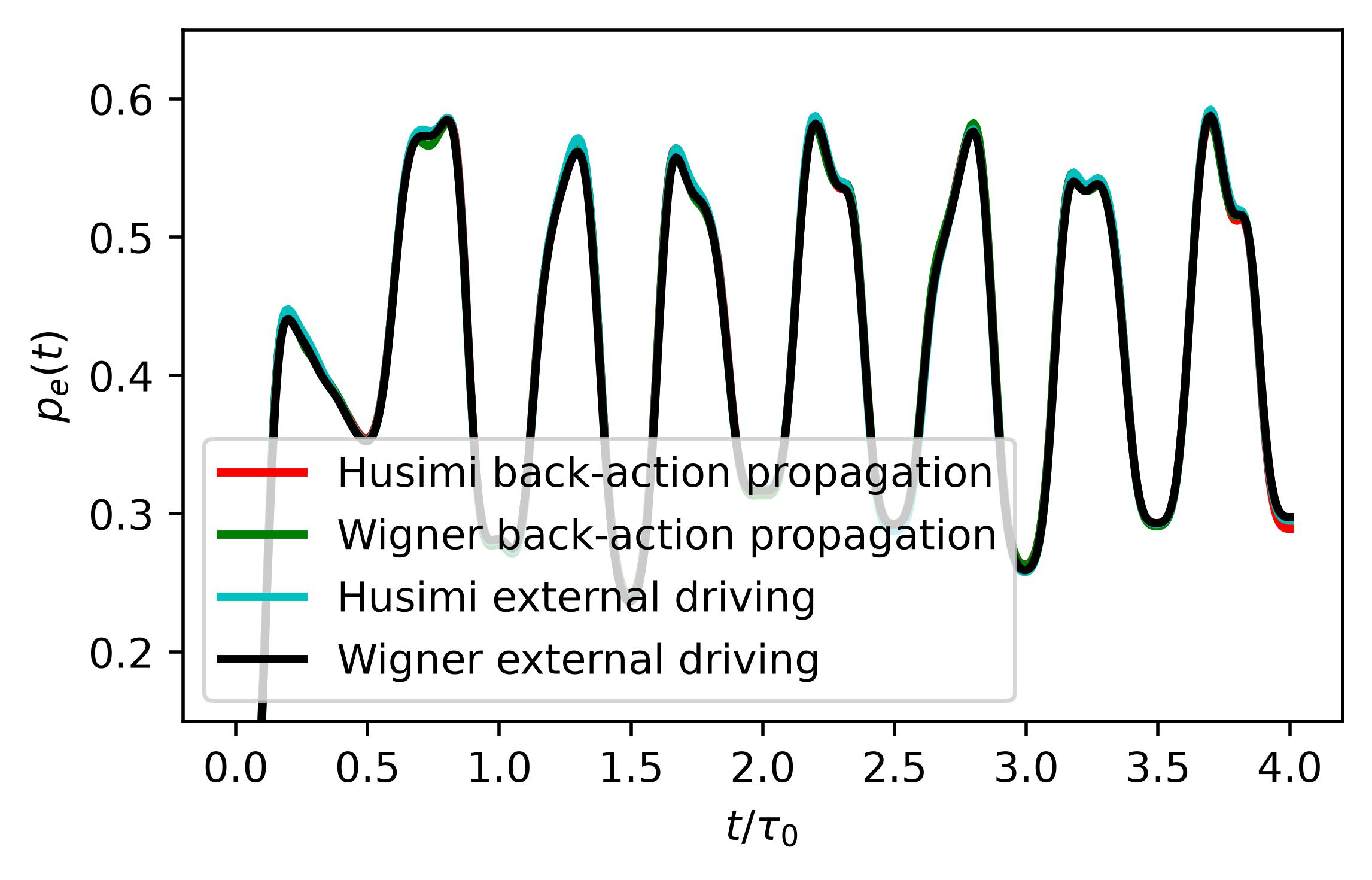}
    \caption{Occupation of the excited state over time, $p_e(t)$, employing different semiclassical methods: (i) Husimi sampling with back-action propagation, (ii) Wigner sampling with back-action propagation, (iii) Husimi sampling as external driving, and (iv) Wigner sampling as external driving.}
    \label{fig: semiclassical-methods}
\end{figure*}

\noindent which basically means that the external driver is sampled from a probability distribution at initial time. Now, BSV has a  preculiar shape in phase space when one represents is through the Wigner or the Husimi distribution. They are both Gaussian distributions centered at zero with squeezed-antisqueezed quadrature widths given by $\left\lbrace e^{-\xi}/\sqrt{2}, e^{\xi}/\sqrt{2} \right\rbrace$ and $\left\lbrace \left( 1+ e^{-\xi} \right)/\sqrt{2}, \left( 1+e^{\xi}\right) /\sqrt{2} \right\rbrace$, respectively for Wigner and Husimi distributions. In the limit of $\left| \xi \right| \rightarrow \infty$ (and simultaneously $\lambda \rightarrow 0$) we find that the phase-space field coordinates $\mathcal{E}$ tend for both Wigner and Husimi distributions, to $\left\lbrace 0, \sqrt{2}\Omega_R \right\rbrace$. Therefore, all semiclassical phase-space methods lead to the same numerical result in the infinite-photon limit. We numerically confirmed this result by comparing the excited state population over time, $p_e(t)$, employing the four methods: (i) Husimi sampling with back-action propagation, (ii) Wigner sampling with back-action propagation, (iii) Husimi sampling as external driving, and (iv) Wigner sampling as external driving. The results of such test can be found on Fig.~\ref{fig: semiclassical-methods}. We compare such result with the exact calculation when computing the quantum state of the two-level system, which is completely defined by its excited-state occupation $p_e(t) \equiv \bra{\Psi(t)} (1-\hat{\sigma}_x) \ket{\Psi(t)}$ since the density matrix is exactly given by 

\begin{align}
    \hat{\rho}_{\mathrm{TLS}}(t) =
    \begin{pmatrix}
        1-p_e(t) & 0 \\
        0 & p_e(t)
    \end{pmatrix},
\end{align}

\begin{figure*}[ht]
    \centering
    \includegraphics[width=0.5\linewidth]{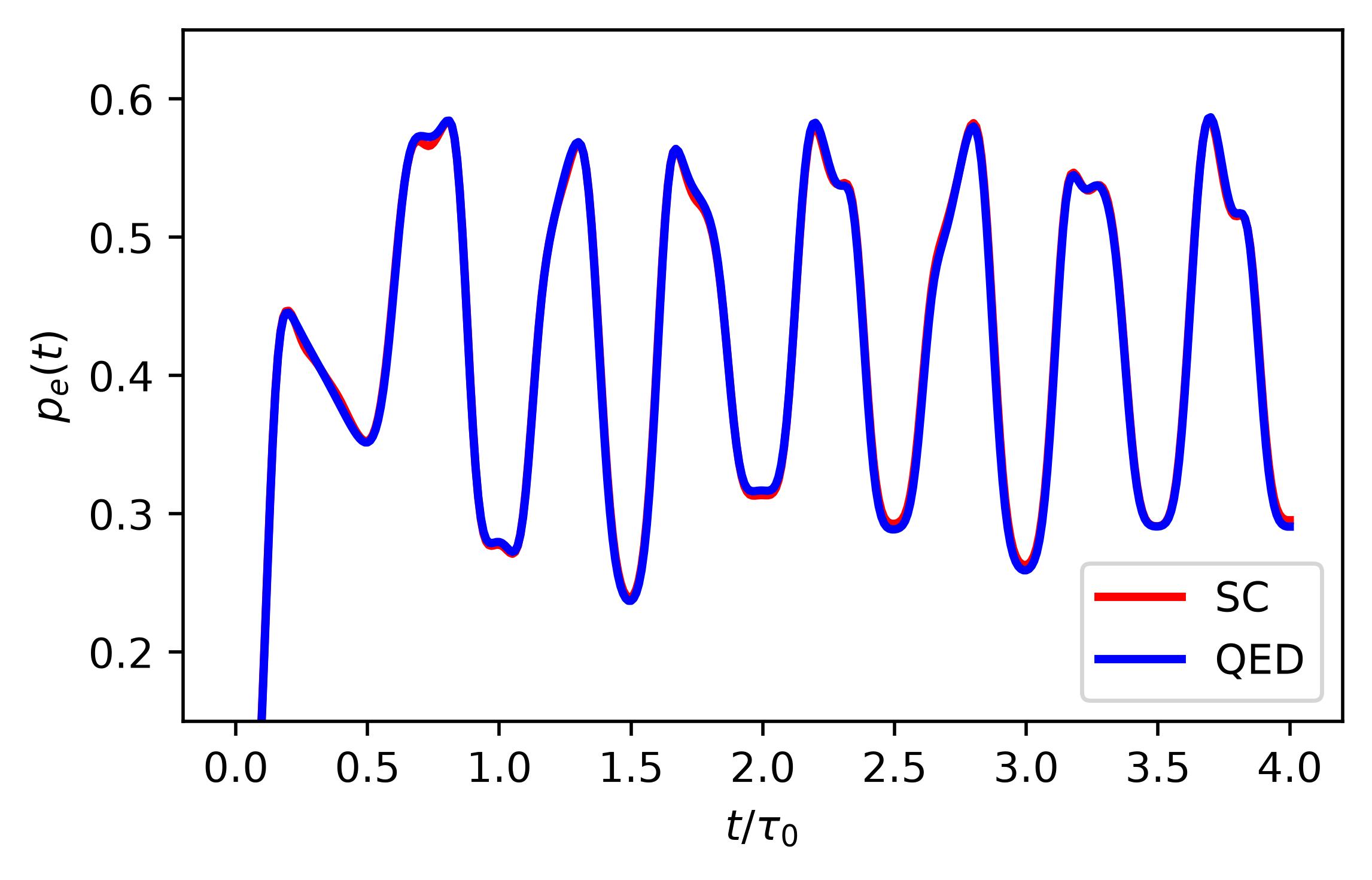}
    \caption{Occupation of the excited state over time, $p_e(t)$, employing exact propagation (QED, blue) and semiclassical propagation (SC, red). Note that the semiclassical propagation leads to the same result regardless of the specific method, as evidenced in Fig.~\ref{fig: semiclassical-methods}.}
    \label{fig: qed-sc-text}
\end{figure*}

\noindent as correlations in the matter state remain zero throughout its evolution due to the symmetry of the BSV driver. Figure~\ref{fig: qed-sc-text} shows $p_e(t)$ employing both the exact calculation and the semiclassical phase-space methods (see SI for a further discussion on the different semiclassical methods). The simulation parameters are a Rabi frequency employed was $\Omega_R/\omega_L = 3.0$ and a resonant condition $\Delta/\omega_L= 1$. The light-matter was checked for values below $\kappa / \omega_L < 10^{-7}$.

\section{Wigner distribution in the squeezed representation}

\noindent To reconstruct the Wigner distribution of the driving mode in the squeezed picture, one must realize that the squeezing transformation in Eq.~\eqref{eq: driving-ham-sq} means an implicit symplectic transformation (rotation and shearing) in the Wigner phase-space: $W^{(s)}({\bf x}_L) = W(S^{-1}(\xi) {\bf x}_L)$ with ${\bf x}_L = (q_L, p_L)^{\mathrm{T}}$  being the phase-space coordinate 

\begin{align}
    S(\xi) = R(\varphi/2)
    \begin{pmatrix}
        e^{-\left| \xi \right|} & 0 \\
        0 & e^{\left| \xi \right|}
    \end{pmatrix}
    R(-\varphi/2)
\end{align}

\begin{align}
    R(\theta) = 
    \begin{pmatrix}
        \cos{\theta} & -\sin{\theta} \\
        \sin{\theta} & \cos{\theta}
    \end{pmatrix}
\end{align}

\noindent Hence, one can always easily recover the Wigner distribution in the original representation $W({\bf x}_L)$ from the Wigner distribution in the squeezed representation $W^{(s)}({\bf x}_L)$ by rotating and rescaling the axis. The Wigner distribution in the squeezed representation is reconstructed via the squeezed-Fock basis representation of its traced density operator $\hat{\rho}^{(s)}_{L}(t) = \mathrm{Tr}_{\mathrm{TLS},\mathrm{cont}} \left\lbrace \ket{\Psi^{(s)}(t)} \bra{\Psi^{(s)}(t)} \right\rbrace$ where $\mathrm{Tr}_{\mathrm{TLS},\mathrm{cont}}$ represents the trace over the two-level system and the mode-continuum subspaces. The Wigner distribution is computed using the Wigner function of the Fock basis $\ket{n}\bra{m}$, $w_{nm} ({\bf x})$, such that:

\begin{align}
    W^{(s)}({\bf x}_L ; t) = \sum_{n,m} \bra{n}_L \hat{\rho}^{(s)}_{L}(t) \ket{m}_L w_{nm}({\bf x}_L).
\end{align}

\noindent Therefore, the Wigner distribution will always be expressed in the squeezed representation. Additionally, the Wigner negative volume $\mathcal{V}_-$ defined as the proportion of the phase space whose Wigner distribution is negative weighted by the distribution itself transforms trivially with the squeezing transformation since it is a symplectic transformation. Hence:

\begin{align}
    \mathcal{V}_{-}(t) = -\int_{\mathcal{D}_-} W(q_L,p_L;t) dq_L dp_L = -\int_{\mathcal{D}^{(s)}_-} W^{(s)}(q_L,p_L;t) dq_L dp_L,
\end{align}

\noindent where $\mathcal{D}_-$ is the region of phase-space where the Wigner distribution is negative and $\mathcal{D}^{(s)}_-$ is the region of the squeezed-phase-space with negative Wigner distribution. The negative volume is hence preserved in the squeezing transformation and we can compute it directly from the calculated Wigner distribution.

\newpage

\section{Complete figure of the third-order correlation}

\newpage

%\section{Wigner distribution movie}

%\href{run:wigner-ev.mp4}{\textbf{Play movie}}

\end{document}